\documentclass[useAMS,usegraphicx,usenatbib]{mn2e}
\usepackage{times}
\usepackage{amsmath}
\usepackage{amssymb}
\usepackage{float}
\usepackage{rotfloat}
\usepackage{graphicx}

\def\mnras{MNRAS}
\def\aj{AJ}
\def\aap{A\&A}
\def\apj{ApJ}
\def\apjl{ApJ}
\def\apjs{ApJS}

\def\pasp{PASP}

\def\procspie{SPIE}

\title[Host galaxies of luminous quasars: population synthesis]
      {Host galaxies of luminous quasars: population synthesis of optical
      off-axis spectra}

\author[Wold et al.]
       {I.\ Wold,$^{\! 1}$
A.\,I.\ Sheinis,$^{\! 1}$
M.\,J.\ Wolf,$^{\! 1}$ and
E.\,J.\ Hooper$^{\! 1}$
\vspace*{1mm}\\
$^1$  Dept of Astronomy, University of Wisconsin-Madison, 475 N.\ Charter St.,
     Madison, WI\,53706, USA}

\date{Accepted ... ; Received ... ; in original form ...}
\pagerange{000--000}

\begin{document}
\maketitle
\begin{abstract}
There is increasing evidence of a connection between AGN activity
and galaxy evolution. To obtain further insight into this potentially
important evolutionary phase, we analyse the properties of quasar
host galaxies. In this paper, we present a population synthesis modeling
technique for off-axis spectra, the results of which constrain host
colour and the stellar ages of luminous quasars ($M_{V}(nuc)<-23$).
Our technique is similar to well established quiescent-galaxy models,
modified to accommodate scattered nuclear light (a combination of
atmospheric, instrumental and host galaxy scattered light) observed
off axis. In our model, subtraction of residual scattered quasar light
is performed, while simultaneously modeling the constituent stellar
populations of the host galaxy. The reliability of this technique
is tested via a Monte-Carlo routine in which the correspondence between
synthetic spectra with known parameters and the model output is determined.
Application of this model to a preliminary sample of 10 objects is
presented and compared to previous studies. Spectroscopic data was
obtained via long-slit and integral-field unit observations on the
Keck and WIYN telescopes. We confirm that elliptical quasar hosts
are distinguishable (bluer) from inactive ellipticals in rest frame
\textit{B-V} colour. Additionally, we note a trend for radio luminous
($L_{5GHz}\gtrsim10^{40}\: erg\: s^{-1}$) quasars to be located in
redder host galaxies in comparison to their less luminous radio counterparts.
While the host colour and age of our radio luminous sample is in close
proximity to the green valley, our radio faint sample is consistent
with quiescent star-forming galaxies. However, further observations
are needed to confirm these results. Finally, we discuss future applications
for our technique on a larger sample of objects being obtained via
SALT and WIYN telescope observing campaigns. 
\end{abstract}

\section{Introduction}

It is well established that active galactic nucleus (AGN) activity
is due to accretion onto super massive black holes (SMBHs). Additionally,
SMBHs are found to be common, existing in most, if not all, massive
galaxies. Observationally, the SMBH mass is found to be tightly correlated
to the bulge velocity dispersion raised to the fourth power \citep{gebhardt00}.
However, the bulge extends well outside the gravitational influence
of the SMBH. This correlation has led to the hypothesis that the bulge
and SMBH co-evolve \citep{kormendy00}. Additionally, models of galaxy
evolution must include regulation mechanisms to adjust and quench
star formation in massive galaxies after early times to prevent the
over production of bright galaxies \citep{benson03}. 

Evidence pointing toward the co-evolution of SMBHs / bulges and the
need for regulation mechanisms has led to models which utilize AGN
feedback to quench star formation. An example of a model of relevance
to our studies is the hydrodynamic simulations of \citet{hopkins06}.
In this model, gas rich mergers induce gas inflow triggering both
star formation and quasar activity. Some small percentage of AGN energy
output is converted to thermal energy, assisting in the stoppage of
further star formation and accretion. Simulations, which feature self-regulated
SMBH growth, have been successful in reproducing various observables,
such as the bimodal colour-magnitude distribution \citep{cattaneo06}.
However these models must operate on a very large range of scales
($\mu pc$ - $Mpc$) and must incorporate many physical processes.
Therefore it is desirable to further constrain these models observationally.
The goal of this work is to develop an off-axis method to provide
stellar age and host colour constraints during the quasar phase.

\begin{table*}
\noindent \centering{}\begin{tabular}{cccccc}
Object name & \multicolumn{1}{c}{redshift} & $M_{V}(nuc)$ & $M_{V}(host)$ & \begin{tabular}{c}
$log(L_{5GHz})$\tabularnewline
erg s$^{-1}$\tabularnewline
\end{tabular} & \begin{tabular}{c}
Host\tabularnewline
Morphology\tabularnewline
\end{tabular}\tabularnewline
\hline 
3C 273  & 0.1583 & -26.7 & -23.2 & 44.1 & E\tabularnewline
4C 31.63  & 0.2950 & -25.1 & -25.1 & 43.3 & E\tabularnewline
PKS 1302-102  & 0.2784 & -25.9 & -22.9 & 43.0 & E\tabularnewline
PKS 0736+017  & 0.191 & -23.2 & -22.6 & 43.0 & E\tabularnewline
PKS 2135-147  & 0.2003 & -24.7 & -22.4 & 42.9  & E\tabularnewline
PKS 2349-014  & 0.1740 & -24.5 & -23.2 & 42.5  & E\tabularnewline
PG 1309+355  & 0.1840 & -24.4 & -22.8 & 41.3  & S\tabularnewline
PHL 909  & 0.171 & -24.1 & -22.2 & 40.0  & E\tabularnewline
PG 0052+251  & 0.1550 & -24.1 & -22.5 & 39.4  & S\tabularnewline
PG 1444+407  & 0.2673 & -25.3 & -22.7 & 39.2 & S\tabularnewline
\end{tabular}\caption{Test sample properties. $M_{V}(nuc)$ is the V-band absolute magnitude
of the quasar. $M_{V}(host)$ is the V-band absolute magnitude of
the host galaxies. Both of these quantities are derived from \citet{bahcall97}
when possible, see \citet{wolf08} for details. Radio luminosity is
estimated utilizing the NASA/IPAC Extragalactic Database and assuming
spectral index of -0.5. Host morphology is obtained via \citet*{hamilton02}.
'E' denotes elliptical and 'S' denotes spiral.}

\end{table*}

Unfortunately, studies of the underlying stellar populations are hampered
by the overwhelming emission from the central quasar. Consequentially,
progress has been slow and, in some cases controversial, in the struggle
to place observational constraints on quasar host galaxies. One group
of collaborators \citep{mclure99,hughes00,nolan01,dunlop03} believes
that local quasar ($M_{V}(nuc)<-23.5$) hosts are predominantly normal
massive elliptical galaxies. However, \textit{Hubble Space Telescope}
(\textit{HST}) morphology \citep*{hooper97,bennert08}, multicolour
imaging \citep*{jahnke04}, and spectral studies \citep{canalizo01,miller03}
have found evidence inconsistent with a population of normal relaxed
ellipticals. In fact, an on-axis spectroscopic study probing the inner
region of nearby quasar hosts found young Sc-like stellar populations
in half of their sample \citep{letawe07}. The discrepancies in these
studies may be due to relatively small sample sizes, systematics,
and the different radii probed \citep{lacy06}. Additional study is
required to develop a clear understanding of stellar properties of
nearby luminous quasar hosts. 

In \citet{wolf08} we presented velocity dispersion measurements of
a sample of nearby luminous quasar host galaxies. Future work will
examine the relation between SMBH mass and velocity dispersion (Sheinis
et al. 2010 in preparation). In this paper, we develop a method to
measure the stellar age and colour of nearby quasar host galaxies
at off-axis radii of 9 to 15 kpc. We observe off-axis spectra to minimize
the observed quasar emission, thereby maximizing signal-to-noise.
However even in off-axis observations, described in \S2, scattered
light from the central source can still contribute significantly to
the observed light. Our technique of removing this scattered light
is similar to previous studies \citep*{boroson85,miller03} and described
in detail in \S3.1. The modelling of the host spectra (\S 3.1) and
the error estimation technique (\S 3.2) are based on quiescent galaxy
models (\citealt{tremonti03}; \citealt[][hereafter CF05]{cid2005}).
Application of our model to our preliminary sample is found in \S4.
A comparison to recent spectroscopic and photometric studies is found
in \S5. Summary and conclusions are presented in \S6. Appendix A
contains comments on individual objects. Throughout this paper a cosmology
of $H_{0}=70\: km\: s^{-1}\: Mpc^{-1}$ , $\Omega_{M}=0.3$, and $\Omega_{\Lambda}=0.7$
is adopted.

\section{Observations }

\subsection{Sample }

Our ongoing spectroscopic observing campaign has resulted in a current
sample of 28 nearby luminous quasars. This sample consists of objects
previously imaged by \citet{bahcall97}, with the addition of a few
objects from \citet{dunlop03} and \citet*{guyon06}. We demonstrate
the capabilities of our spectral synthesis model on a subset of 10
objects. These objects have spectral data that display sufficient
signal-to-noise and stellar continuum to constrain the stellar properties
via the method discussed below. Additionally, this subset of ten objects
has known stellar velocity dispersions \citep{wolf08}, eliminating
a potential free parameter of the model. This relatively small sample
of quasars is not representative of the local quasar population, which
consists of $\sim$10\% radio loud quasars. Our sample contains six
radio loud and four radio quiet objects. We define an object as radio
loud by the criteria established by \citet{kellermann94}, $L_{5\: GHz}>10^{25}W\: Hz^{-1}$,
which is roughly $10^{41.5}erg\: s^{-1}$ for our adopted cosmology%
\footnote{\citet{kellermann94} observationally determined the luminosity threshold
assuming a cosmology of $H_{0}=70\: km\: s^{-1}\: Mpc^{-1}$ , $q_{0}=1/2$.
We solve for the observed flux at $z=0.2$ and then convert to our
adopted cosmology. The specific luminosity is multiplied by the observed
frequency to obtain the quoted 5 GHz luminosity. %
}. All quasars are local $z<0.3$ and luminous $M_{V}<-23$. The properties
of our test sample are summarized in Table 1. 

\begin{flushleft}
\begin{figure*}
\begin{centering}
\includegraphics[width=17cm]{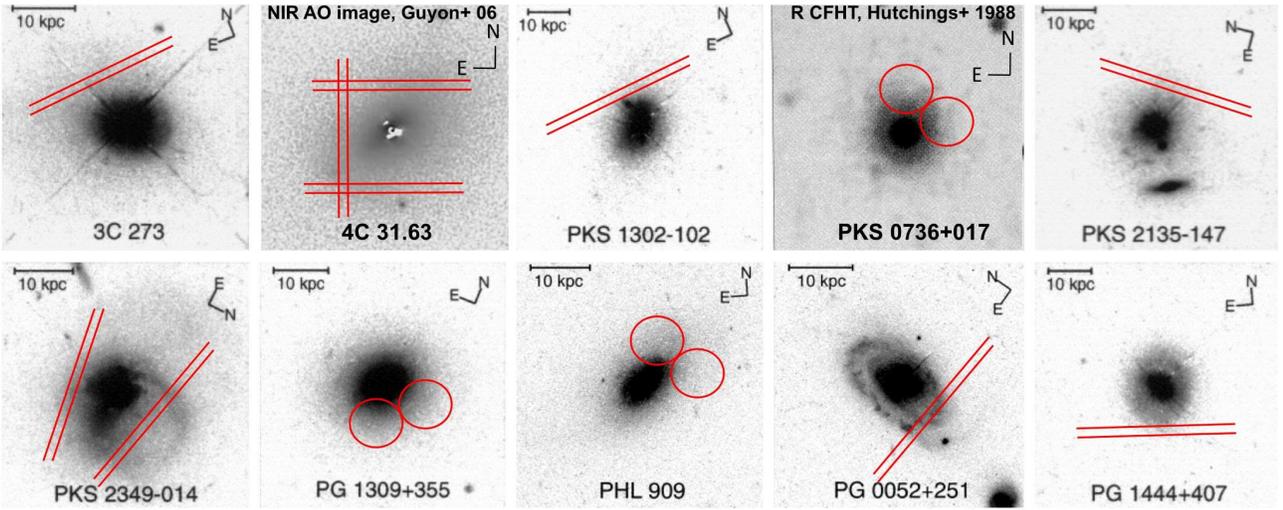}
\par\end{centering}

\caption{Sample of observed quasar hosts with approximate off-axis slit and
fiber positions indicated in red. The slit lengths are much larger
than indicated, extending approximately 7$'$ projected on the sky.
For WIYN objects, only the off-axis fibers used for the host galaxy
spectrum are shown. Objects are listed from left to right, top to
bottom by most radio bright to least. All images are from \textit{HST}
\citep{bahcall97} (23$''$$\times$23$''$) with the exception of
4C 31.63 (10$''$$\times$10$''$) \citep{guyon06} and PKS 0736+017
(30$''$$\times$30$''$) \citep*{hutchings88}. Where indicated,
the 10 kpc scale is computed for $ $$\Omega_{0}=1.0$ and $H_{0}=100$
km $s^{-1}$ $Mpc^{-1}$ \citep{bahcall97}. For our adopted cosmology
this scale corresponds to $\sim15.7$ kpc.\textbf{ }}

\end{figure*}

\par\end{flushleft}

\subsection{Spectra}

Data acquisition and reduction is described in detail in \citet{sheinis02},
\citet{miller03}, and \citet{wolf08}. In summary, seven of the ten
objects were observed via the Low Resolution Imaging Spectrograph
\citep{oke1994} on the Keck 10-m telescope. The remaining three objects
were observed using a 82 fiber integral field unit (IFU), SparsePak
\citep{bershady2005} which feeds the Bench Spectrograph on the 3.5-m
WIYN telescope. Short on-axis observations, typically 1-2 minutes
for Keck and 30 minutes for WIYN, and longer off-axis (2$''$-4.5$''$
from the nucleus) observations, typically 0.5 - 4 hours, were obtained
for each object. The approximate off-axis slit and fiber positions
with respect to archival imaging is displayed in Fig. 1. The properties
of the off-axis pointings are summarized in Table 2. Keck objects
have a rest wavelength range of $\sim$3500-7000\AA $\:$at a spectral
resolution of $\Delta\lambda\sim11$\AA. WIYN objects have a rest
wavelength range of $\sim$3500-6000\AA $\:$at a spectral resolution
of $\Delta\lambda\sim5$\AA. All spectra were corrected for Galactic
extinction using the law of \citet*{cardelli1989} and the $A_{V}$
values from \citet*{schlegel1998} as listed in the NASA/IPAC Extragalactic
Database%
\footnote{http://nedwww.ipac.caltech.edu/%
} (NED). 

\begin{table*}
\noindent \centering{}\begin{tabular}{ccccc}
Pointing & \multicolumn{1}{c}{$R_{obs}$(kpc)} & \begin{tabular}{c}
S/N \AA$^{-1}$\tabularnewline
(5500-5700 \AA)\tabularnewline
\end{tabular} & \begin{tabular}{c}
Quasar Scattering\tabularnewline
(3600-5900 \AA)\tabularnewline
\end{tabular} & Telescope\tabularnewline
\hline 
3C 273 ~ 4N & 11.79 & 18.6 & 68.8\% & Keck\tabularnewline
4C 31.63 ~  2N & 8.74 & 11.5 & 50.5\% & Keck\tabularnewline
4C 31.63 ~  2.5E & 10.92 & 9.4 & 34.6\% & Keck\tabularnewline
4C 31.63 ~  3S & 13.11 & 10.1 & 38.7\% & Keck\tabularnewline
PKS 1302-102 ~ 2.3N & 12.79 & 9.3 & 68.4\% & Keck\tabularnewline
PKS 0736+017 ~ 4.5NW & 14.18 & 8.3 & 49.5\% & WIYN\tabularnewline
PKS 2135-147 ~ 3W & 12.47 & 12.3 & 70.8\% & Keck\tabularnewline
PKS 2349-014 ~ 4N & 14.12 & 14.8 & 21.8\% & Keck\tabularnewline
PKS 2349-014 ~ 3S & 9.47 & 18.8 & 25.2\% & Keck\tabularnewline
PG 1309+355 ~ 4.5SW & 13.76 & 13.4 & 41.3\% & WIYN\tabularnewline
PHL 909 ~ 4.5N & 12.97 & 11.7 & 77.8\% & WIYN\tabularnewline
PG 0052+251 ~ 3S & 9.65 & 23.0 & 64.6\% & Keck\tabularnewline
PG 1444+407 ~ 3S & 15.07 & 8.8 & 69.6\% & Keck\tabularnewline
\end{tabular}\caption{Off-axis spectral properties. Pointing designation indicates object
and approximate off-axis position in arcseconds. $R_{obs}$ is the
observed off-axis radius, as reported in \citet{wolf08}. Additionally,
off-axis signal-to-noise, fraction of the observed off-axis spectrum
which consists of scattered light, and telescope are shown. The scattered
quasar light percentage is determined by the model output. }

\end{table*}

\section{Spectral Synthesis}

\subsection{Model}

\begin{figure*}
\centering{}\includegraphics[bb=60bp 20bp 510bp 800bp,clip,angle=90,width=17cm]{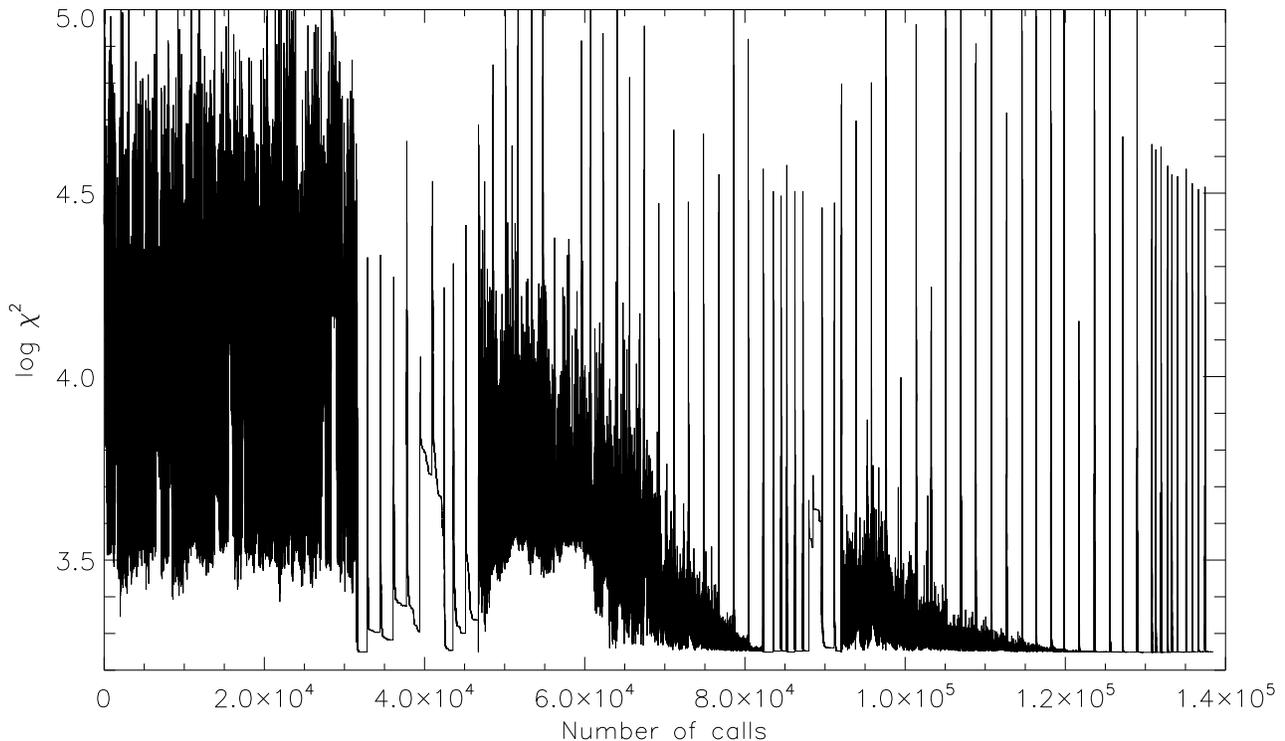}\caption{Parameter space search for minimum $\chi^{2}$ value. The three densely
populated regions consist of 20 simulated annealing runs each. As
the number of calls, or calculations of $\chi^{2}$, advances the
'tempature' is reduced to more restrictive values. These regions are
separated by 10 straight downhill searches. This entire procedure
is performed 20 times independently to determine the model's output.}

\end{figure*}

Our spectral synthesis code is a modified version of the model first
discussed in \citet{miller03}, with the underlying stellar spectrum
now modeled as a weighted summation of 15 instantaneous starbursts
of various post burst ages. We model the off-axis spectrum, which
consists of a stellar component and a scattered quasar component.
Quasar light is scattered into the off-axis line of sight due to atmospheric
seeing effects, as well as a small contribution from the intrinsic
optical point-spread function and astrophysically via dust and gas
in the intervening line of sight. The stellar and scattered quasar
components are modeled simultaneously to achieve the best $\chi^{2}$
fit to the observed spectrum. Each model spectrum is masked in the
same manner as the data for which it is intended (see \S4). A model
spectrum $M_{\lambda}$ is of the form:

\begin{equation}
M_{\lambda}=\left(\sum_{i=1}^{15}x_{i}\, ssp_{i\lambda}\right)\, r_{\lambda}+Q{}_{\lambda}\,\xi_{\lambda}\end{equation}

with the following definitions:
\begin{itemize}
\item $ssp_{i\lambda}$ is the stellar component of the model which consists
of a reduced basis of 15 simple stellar populations (SSPs) from the
\citet{BC03}, BC03, synthesis model, degraded to the instrument resolution,
assuming solar metallicity, Padova-1994 models, and a Chabrier initial
mass function \citep{chabrier03}. Extensive simulations by CF05,
have found this reduced spectral base ($t_{i}$=0.001, 0.00316, 0.00501,
0.01, 0.02512, 0.04, 0.10152, 0.28612, 0.64054, 0.90479, 1.434, 2.5,
5, 11 and 13 Gyr) to reliably produce observed spectral features while
limiting redundancies in quiescent galaxies. Due to the increased
complexity introduced by the scattered light subtraction, only solar
metallicity is considered. Each SSP is normalized such that the area
under the unmasked portions of the spectrum is unity. Normalization
at a single wavelength ($\lambda_{0}$=4020\AA, as prescribed by
CF05) was also investigated. For our sample, the alteration of normalization
convention does not affect the results of our model more than $\sim$1$\sigma$.
$x_{i}$ is the population weighting factor.
\item $r_{\lambda}\equiv10^{-0.4A_{\lambda}}$ is the extinction law used
to model the in situ reddening. $A_{\lambda}(A_{V},\, R_{V})$ is
defined by the Galactic law of \citet{cardelli1989} assuming the
mean value of $R_{V}\:(3.1)$ in the diffuse ISM. A single extinction
is assumed for the entire stellar component. 
\item $Q_{\lambda}$ is the on-axis observed spectrum of the quasar normalized
such that the area under the unmasked portions of the spectrum is
unity. $\xi_{\lambda}\equiv a_{1}+a_{2}\lambda+a_{3}\lambda^{2}+a_{4}\lambda^{3}$
is the scattering efficiency curve, which modifies the observed quasar
spectrum with the goal of modeling the scattered quasar light observed
off axis. We have shown that this model component is necessary to
reliably remove scattered quasar light \citep{sheinis02}. The model
assumes no host galaxy contamination to the on-axis spectrum. The
dominance of the high luminosity quasar ($M_{V}<-23$) on-axis makes
this assumption acceptable. 
\end{itemize}
\begin{figure*}
\noindent \begin{centering}
\includegraphics[bb=60bp 425bp 725bp 540bp,clip,angle=180,width=17cm]{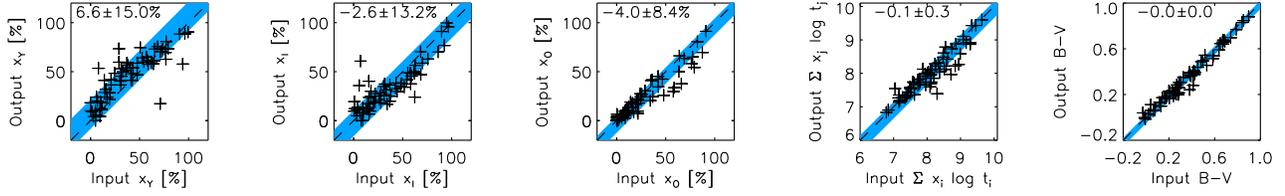}\caption{Monte-Carlo simulation results for ideal noiseless off-axis spectra
with 60\% scattered quasar light contamination. 65 synthetic spectra,
$S_{\lambda}$, are analysed by the model to determine the reliability
of the results. For each $S_{\lambda}$, the known young, intermediate
and old population percentage and the known $\left\langle log\: t_{\star}\right\rangle _{L}$
and rest frame \textit{B-V} input is compared to the model output
as represented by 65 plus signs in each box. Ideally, all points would
fall on the diagonal, denoted by a dashed line. The mean offset from
the ideal and the standard deviation of the model output are displayed
numerically in the upper left corner and graphically by the blue strip. }

\par\end{centering}

\end{figure*}

\begin{flushleft}
Finally, the individual pieces are combined as prescribed by equation
(1) and the resulting model spectrum, $M_{\lambda}$, is normalized,
dividing by a single constant, such that the area under the unmasked
portions of the spectrum is unity (the normalized model is denoted
$M_{\lambda n}$). 
\par\end{flushleft}

\subsubsection{Parameter range}

Our model contains 20 free parameters: (1) an extinction parameter
$A_{V}$, (15) SSP weighting factors $x_{i}$, (4) the scattering
efficiency curve coefficients, $a_{[1:4]}$. The range of values to
explore with the model is constrained in the case of $x_{i}$ to be
0.0 to 1.0, where $\Sigma x_{i}=1.0$. The range of values for $A_{V}$
and $a_{[1:4]}$ were found in an iterative fashion. First the model
was applied to the entire dataset with parameter ranges informed from
the work of \citet{sheinis02} and \citet{miller03}. The range was
then increased as necessary to ensure the best $\chi^{2}$ fit was
not determined by the searchable parameter space. For example, the
highest model $A_{V}$ output obtained is 0.4; therefore, the model
samples $A_{V}$ from 0.0 to 1.0. We have tested the sensitivity of
the model output to these boundary conditions by doubling the range
and re-running the model on the observed data. No significant change
in results is found.

\subsubsection{Parameter search}

Having constructed a model, we then devised a reliable means to find
the minimum chi-squared fit. The $\chi^{2}$ fitting function is defined
by:

\begin{equation}
\chi^{2}={{\displaystyle \sum}\atop {\scriptstyle \lambda}}{\displaystyle \,\left(\frac{O_{\lambda n}-M_{\lambda n}}{\sigma_{\lambda}}\right)^{2}}\end{equation}

$O_{\lambda n}$ is the observed off-axis spectrum, normalized in
the same manner as $M_{\lambda n}$, and $\sigma_{\lambda}$ is the
observed spectral noise. The noise spectrum is generated by measuring
the standard deviation of $O_{\lambda n}$ at three to four locations
devoid of prominent spectral features. These measurements are then
extrapolated across the wavelength range of $O_{\lambda n}$ by a
low order fitting polynomial. 

The starting point in the parameter search is defined in the following
manner. The initial stellar component is set to be 100\% 5Gyr SSP.
The reddening is randomly selected, constrained by its range. The
scattering coefficients are also randomly selected but then scaled
so that scattered quasar light comprises 60\% of the off-axis model
spectrum. Parameters $a_{[1:4]}$ are further constrained by rejecting
any solutions which drive the resulting off-axis quasar flux, $Q{}_{\lambda}\,\xi_{\lambda}$
, below zero. 

Simulated annealing optimization \citep{press1992} is utilized to
search the 20-D parameter space. Simulated annealing evaluates the
merit function, $\chi^{2}$ in this case, and travels downhill in
merit space to find a minimum within predefined tolerances. To help
avoid local minima, a user defined parameter, traditionally called
'temperature', is set to determine the probability of travelling uphill
in merit space. The higher the temperature the more probability of
traveling uphill in the merit space. A zero value for temperature
corresponds to a straight downhill search. 

The utilization of this optimization routine, described below, is
somewhat of an art. The temperature value must be decreased slowly
so that the merit space can be adequately sampled, yet fast enough
to arrive at a timely solution. After extensive experimentation the
following implementation has been found to produce favorable results.
First, twenty iterations of simulated annealing are performed with
linearly decreasing temperature. If an iteration fails to find a minimum
$\chi^{2}$ value within a factor of two of the overall minimum then
the next iteration is reset to the overall minimum; otherwise the
next iteration starts where the previous iteration left off. At the
completion of the simulated annealing runs, the ten best distinct
solutions are used as input for straight downhill (temperature = 0)
runs. The current global minimum is then used as the input to the
next twenty iterations of simulated annealing. This procedure is carried
out for a total of three loops; within each loop the temperature parameter
is set to more restrictive values. An example of this parameter space
search is shown in Fig. 2 for a typical $\chi^{2}$ minimization.
Overall approximately 150,000 points are sampled in the parameter
space. This entire procedure is performed 20 times independently.
The best $\chi^{2}$ value of these runs, which consists of approximately
$3\times10^{6}$ calculations of $\chi^{2}$, is defined as the model
output.

\subsubsection{Quantities of interest}

Guided by the analysis of CF05, we limit the scope of our project
to a coarse, but well recovered, description of the stellar populations.
Like CF05, we only attempt recovery of \textquoteleft{}young\textquoteright{}
(t $<$ 100 Myr), \textquoteleft{}intermediate\textquoteright{} (100
Myr \ensuremath{\le} t \ensuremath{\le} 1 Gyr), and \textquoteleft{}old\textquoteright{}
(t $>$ 1 Gyr) stellar contributions ($x_{Y}$, $x_{I}$, and $x_{O}$,
respectively). Additionally, we report the mean light-weighted stellar
age,

\begin{equation}
\left\langle log\: t_{\star}\right\rangle _{L}=\sum_{i=1}^{N_{\star}=15}x_{i}\, log\: t_{i}\end{equation}

\noindent \begin{flushleft}
and the rest frame B-V colour as derived directly from the listed
BC03 SSP values. These quantities have been studied extensively in
quiescent galaxy samples and to a lesser extent in AGN host galaxy
samples. Thus, we may assess the consistency of our results to previous
AGN host studies. Furthermore, we may compare these properties of
quasar host galaxies to quiescent galaxy samples. Analysis of other
quantities of interest (see CF05) are deferred to future studies.
\par\end{flushleft}

\subsection{Monte-Carlo Simulations}

The error in our method is estimated via a Monte-Carlo routine. Synthetic
spectra, $S_{\lambda}$, are generated with known parameters and then
degraded by noise.  The same procedure used for the observed data
is applied to determine the agreement between the known input and
model output. In other words, $S_{\lambda n}$ replaces $O_{\lambda n}$
in the fitting routine. $S_{\lambda}$ are constructed in the same
manner as $M_{\lambda}$ with the exception of metallicity, SSP base
and added noise. Because the observed galaxies will likely include
non-solar metallicities, even though the model grid does not, the
metallicity is allowed to deviate from solar, and is randomly selected
from $Z=0.2,\,1$ and $2.5$ $Z_{\odot}$. The SSP base is extended
to include all 221 BC03 instantaneous bursts assuming a Chabrier initial
mass function. A synthetic spectrum is of the form:

\begin{equation}
S_{\lambda}=\left(\sum_{j=1}^{3\times221}x_{j}\, ssp_{j\lambda}\right)\, r_{\lambda}+Q{}_{\lambda}\,\xi_{\lambda}\end{equation}

Poison noise is added until the synthetic spectra obtain the corresponding
observed noise as measured in the 5500-5700\AA $\:$ spectral window.
The 5500-5700\AA $\:$ spectral range was selected because there
are no prominent emission or absorption features; additionally, all
observed spectra include this range. The value of $A_{V}$ and $a_{[1:4]}$
are randomly selected, constrained by the applicable range. Parameters
$a_{[1:4]}$ are further constrained by rejecting any solutions which
drive the resulting off-axis quasar flux, $Q{}_{\lambda}\,\xi_{\lambda}$,
below zero. The scattered quasar flux is then scaled to the contamination
percentage of interest. The value of \textbf{$x_{j}$} is randomly
selected to uniformly sample%
\footnote{To uniformly sample each stellar age bin the following procedure is
employed: The first of 65 synthetic spectra, $S_{\lambda}$, is randomly
assigned a $x_{O}$ value between 0-100\%. $x_{I}$ and $x_{Y}$ are
then randomly assigned the remaining flux. The second (third) $S_{\lambda}$
is constructed similarly, except $x_{I}$ ($x_{Y}$) is first assigned
a value between 0-100\%. This cycle is repeated beginning with the
forth $S_{\lambda}$. To further illustrate the procedure let us consider
the construction of the first $S_{\lambda}$. Suppose $x_{O}$ was
assigned a value of 72\%. The program populates $x_{O}$ by randomly
selecting a SSP and a metallicity and then randomly assigning a percentage
of the total $x_{O}$ value. This process continues until $x_{O}$=72\%
to within $\pm$0.005\%. $x_{I}$ and $x_{Y}$ are then populated
in the same manner. Thus, the stellar components of the synthetic
spectra are constructed.%
} each stellar age bin (young, intermediate, and old). %
\begin{figure*}
\begin{centering}
\includegraphics[bb=40bp 120bp 383bp 730bp,clip,angle=90,width=14cm]{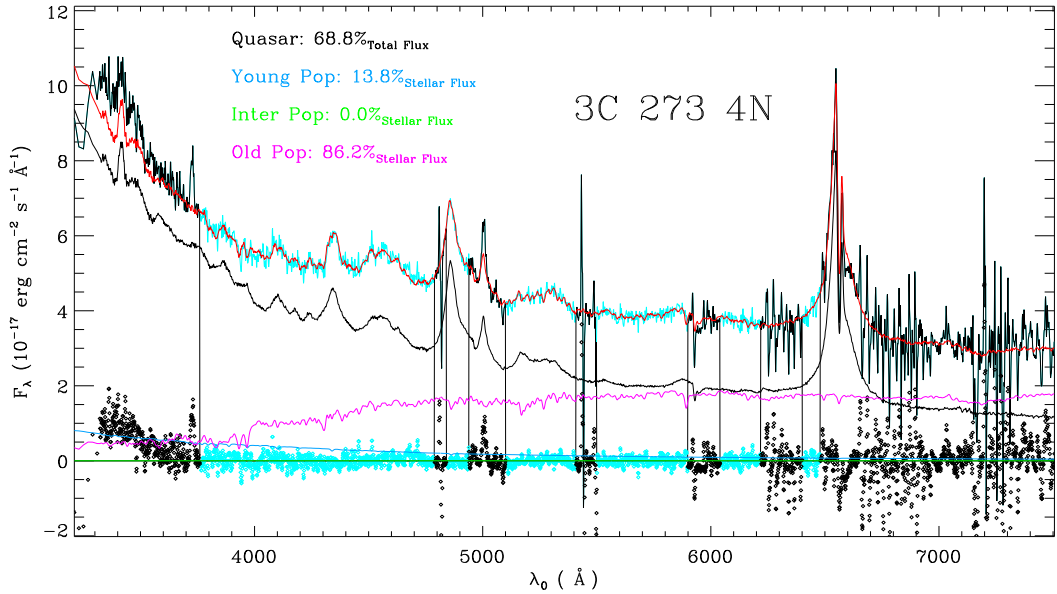}
\par\end{centering}

\caption{\label{fig3} Model fit to the 3C 273 4N pointing. The observed off-axis
rest frame spectrum is highlighted in cyan, with masked regions indicated
in black and by vertical lines extending to zero flux; masked regions
that are narrow, {[}OIII{]} at 5007\AA ~for example, are most easily
seen by looking at the residuals at the bottom of the plot. The model
output is over-plotted in red. The residual is shown with diamonds.
Model components are also shown. The scattered light component is
indicated by the thin black line and comprises 68.8\% of the total
observed flux. The stellar component is also shown, broken up into
young (blue), intermediate (green), and old (magenta) bins. In this
example, only the old stellar component, which comprises 86.2\% of
the stellar flux, is clearly visible. Results for all other pointings
are found in Appendix A, Fig. A1. }

\end{figure*}
Thus, synthetic off-axis spectra are generated with generalized metallicity
and SSP base. Sixty-five unique $S_{\lambda}$ are constructed to
test the ability of the model to recover known input parameters. We
will use this Monte-Carlo routine first to test the associated uncertainties
in a noiseless test case (\S3.2.1) and the reliability to recover
the scattered nuclear light (\S3.2.2), before we will apply it to
our actual data to determine the errors and biases in the recovered
model parameters (\S4).

\subsubsection{Model performance given noiseless spectra}

A concern regarding the use of our multi-component model is inherent
degeneracies. For instance, can a young stellar population mimic scattered
quasar light? The legitimacy of the model relies on the subtle differences
between the various components. The reddening function, which only
operates on the stellar component, should be distinct from the scattering
function which operates on the observed on-axis spectrum. The absorption
features in the stellar component generally do not match quasar absorption
lines. The scattering function, applied to the quasar spectrum, uniquely
has broad-line components. However, can quantities of interest be
reliably recovered given the many possible interactions between model
components?

To test the ideal performance of the model we conducted noiseless
Monte-Carlo simulations. For this purpose, the Sloan Digital Sky Survey
(SDSS) composite quasar spectrum \citep{vandenberk01} was utilized
as the on-axis quasar spectrum, $Q{}_{\lambda}$. The scattered quasar
light was scaled to a value of 60\%. The results of this test are
shown in Fig. 3. Referring to the figure, input versus output values
of $x_{Y}$, $x_{I}$, $x_{O}$, $\left\langle log\: t_{\star}\right\rangle _{L}$
and \textit{B-V} are shown for 65 synthetic spectra. Ideally, all
points would fall on the diagonal, denoted by a dashed line (where
input equals output). The mean offset from the ideal, mean(output-input),
plus the standard deviation of the model output, $\pm$stdev(output-input),
are displayed numerically in the upper left corner and graphically
by the blue strip. The fitting routine reliably recovers $\left\langle log\: t_{\star}\right\rangle _{L}$
and \textit{B-V }in the ideal case of noiseless spectra. The values
of $x_{Y}$, $x_{I}$ and $x_{O}$ are less well constrained, with
systematic offsets of up to 6\% and one sigma deviations of up to
13.4\%.

\subsubsection{Scattered light verification tests}

Additional Monte-Carlo tests were conducted to check for consistency
in the model. Inputting simulated off-axis spectra consisting of 100\%
scattered quasar light and a signal-to-noise of 10, the algorithm
reliably assigned no flux to the stellar component (65 iterations
resulted in a mean result of 99.92$\pm$0.01\% scattered quasar light).
Inputting 100\% stellar component (S/N=10), the algorithm reliably
assigned no flux to the scattered light component (65 iterations resulted
in a mean result of 2.32$\pm$1.59\% scattered quasar light). Finally,
we tested the performance of the model when constrained to have a
steep blue scattered quasar component, as observed in the off-axis
spectrum of PHL 909. The error estimates were not significantly affected
by this alteration; for details see Appendix A (PHL 909).

\subsubsection{Limitations}

\begin{table*}
\noindent \centering{}\begin{tabular}{cccccc}
Pointing & \multicolumn{1}{c}{\begin{tabular}{c||c||c}
\multicolumn{3}{c}{Young stell}\tabularnewline
\multicolumn{3}{c}{pop - \% flux}\tabularnewline
\end{tabular}} & \begin{tabular}{c||c||c}
\multicolumn{3}{c}{Inter stell}\tabularnewline
\multicolumn{3}{c}{pop - \% flux}\tabularnewline
\end{tabular} & \begin{tabular}{c||c||c}
\multicolumn{3}{c}{Old stell}\tabularnewline
\multicolumn{3}{c}{pop - \% flux}\tabularnewline
\end{tabular} & $\left\langle log\: t_{\star}\right\rangle _{L}$ & \textit{B - V}\tabularnewline
\hline 
3C 273 ~ 4N & 13.8 (-3.0$\pm$16.9) & 0.0 (+0.3$\pm$19.1) & 86.2 (+2.6$\pm$13.8) & 9.40 (-0.01$\pm$0.24) & 0.77 (-0.00$\pm$0.05)\tabularnewline
4C 31.63 ~ 2N & 15.7 (-1.9$\pm$16.2) & 0.0 (-1.8$\pm$16.5) & 84.3 (+3.7$\pm$9.6) & 9.00 (+0.06$\pm$0.28) & 0.69 (+0.01$\pm$0.06)\tabularnewline
4C 31.63 ~ 2.5E & 18.3 (-1.1$\pm$16.4) & 0.5 (-2.3$\pm$17.7) & 81.3 (+3.4$\pm$11.8) & 9.18 (+0.00$\pm$0.32) & 0.74 (-0.00$\pm$0.05)\tabularnewline
4C 31.63 ~ 3S$^{\Delta}$ & 0.0 (-2.1$\pm$17.0) & 25.3 (-1.0$\pm$18.2) & 74.7 (+3.1$\pm$9.9) & 9.73 (+0.03$\pm$0.26) & 0.83 (+0.01$\pm$0.05)\tabularnewline
PKS 1302-102 ~ 2.3N & 0.0 (-5.6$\pm$27.0) & 12.4 (+4.3$\pm$26.7) & 87.6 (+1.3$\pm$14.8) & 9.34 (+0.09$\pm$0.49) & 0.76 (+0.02$\pm$0.12)\tabularnewline
PKS 0736+017 ~ 4.5NW & 0.0 (+4.2$\pm$21.3) & 0.0 (-5.7$\pm$23.2) & 100.0 (+1.5$\pm$12.4) & 9.56 (-0.05$\pm$0.32) & 0.83 (-0.01$\pm$0.07)\tabularnewline
PKS 2135-147 ~ 3W & 0.0 (-2.0$\pm$22.6) & 0.0 (-0.6$\pm$23.7) & 100.0 (+2.6$\pm$12.0) & 9.52 (+0.01$\pm$0.34) & 0.82 (-0.00$\pm$0.07)\tabularnewline
$ $PKS 2349-014 ~ 4N & 6.3 (-2.5$\pm$11.4) & 0.8 (+0.0$\pm$11.3) & 92.9 (+2.5$\pm$9.6) & 9.52 (+0.01$\pm$0.20) & 0.79 (+0.00$\pm$0.04)\tabularnewline
PKS 2349-014 ~ 3S & 11.1 (-4.7$\pm$12.7) & 15.3 (+0.8$\pm$13.2) & 73.6 (+3.9$\pm$8.8) & 9.54 (+0.06$\pm$0.21) & 0.78 (+0.00$\pm$0.03)\tabularnewline
PG 1309+355 ~ 4.5SW & 11.7 (-2.1$\pm$14.4) & 9.5 (+1.9$\pm$16.1) & 78.9 (+0.2$\pm$10.7) & 9.33 (+0.02$\pm$0.26) & 0.76 (+0.00$\pm$0.06)\tabularnewline
$ $PHL 909 ~ 4.5N & 0.0 (-8.3$\pm$18.4) & 43.8 (+4.9$\pm$20.0) & 56.2 (+3.4$\pm$12.4) & 9.45 (+0.14$\pm$0.45) & 0.73 (+0.02$\pm$0.10)\tabularnewline
PG 0052+251 ~ 3S & 43.3 (-8.0$\pm$19.2) & 16.5 (+4.5$\pm$20.9) & 40.1 (+3.6 $\pm$11.5) & 8.09 (+0.09$\pm$0.34) & 0.32 (+0.03$\pm$0.05)\tabularnewline
PG 1444+407 ~ 3S & 9.5 (-7.8$\pm$23.8) & 58.2 (+2.6$\pm$27.6) & 32.2 (+5.2$\pm$17.7) & 8.86 (+0.15$\pm$0.42) & 0.52 (+0.03$\pm$0.09)\tabularnewline
\end{tabular}\caption{Modeled off-axis stellar populations, $\left\langle log\: t_{\star}\right\rangle _{L}$
and rest frame colour are shown. Host stellar population offset and
$\pm1\sigma$ error are shown in parentheses as estimated with Monte-Carlo
simulations. $\Delta$ denotes the observation with non-concurrent
on and off-axis observations.}

\end{table*}
\begin{figure*}
\begin{centering}
\includegraphics[bb=60bp 60bp 170bp 725bp,clip,angle=90,width=17cm]{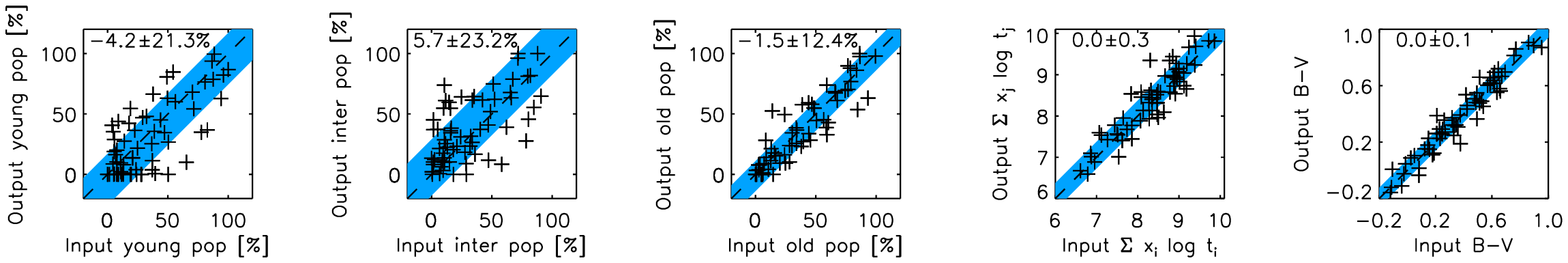}
\par\end{centering}

\begin{centering}
\includegraphics[bb=60bp 60bp 170bp 725bp,clip,angle=90,width=17cm]{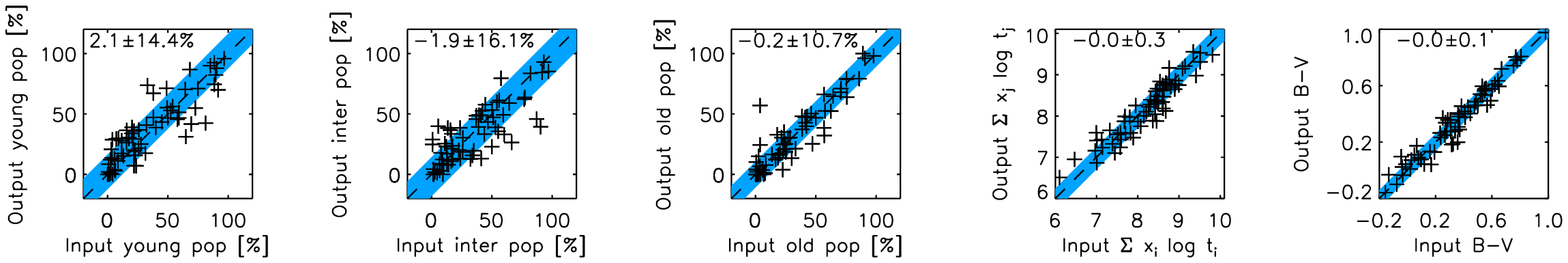}
\par\end{centering}

\centering{}\caption{\textbf{Top row}: Monte-Carlo simulation for PKS 0736+017 (off-axis
pointing of most limiting signal-to-noise). \textbf{Bottom row}: Monte-Carlo
simulation for PG 1309+355 (off-axis pointing of $\sim$average signal-to-noise
and quasar contamination). 65 synthetic spectra are analysed by the
model to determine the reliability of the results. For each $S_{\lambda}$,
the known young, intermediate and old population percentage and the
known $\left\langle log\: t_{\star}\right\rangle _{L}$ and rest frame
\textit{B-V} input is compared to the model output as represented
by 65 plus signs in each box. Ideally, all points would fall on the
diagonal, denoted by a dashed line. The mean offset from the ideal
and the standard deviation of the model output are displayed numerically
in the upper left corner and graphically by the blue strip. }

\end{figure*}

Errors associated with the: 1) population synthesis model (BC03) (template
mis-match), 2) inadequacy of the scattering efficiency curve, 3) non-Poisson
noise and 4) the application of a single simplistic extinction law
are not represented in our Monte-Carlo simulations .\textbf{ }Despite
these deficiencies, similarly constructed quiescent models have demonstrated
consistency when compared to competing techniques (e.g. see CF05).
In \S 5 we address this issue by comparing our results to recent
studies of quasar host galaxies. Finally, it should be noted that
the error estimates obtained reflect the one sigma performance of
the model. Although currently applied to a relatively small sample,
stronger error statements will be possible with the increase of our
sample. With this in mind, when possible, we consider our objects
as an ensemble.

\section{Application of model and simulations to the test sample}

In this section we apply our model to the test sample to constrain
the stellar age and host colour. We then conduct 10 (one for each
pointing) tailored Monte-Carlo simulations to estimate the reliability
of our results. 

The following caveats should be considered in the application of our
model to the test sample. For each object, concurrent on and off-axis
observations have been obtained, with the exception of the 3S pointing
of 4C31.63 ($\triangle t\sim1.5yr$). The possible variability of
this quasar over this time period is not accounted for by the model.
Line-of-sight stellar motions are not modeled for Keck objects. Our
instrument resolution, $\sigma_{v}\sim300$ km $s^{-1}$, being greater
or of the order of the measured velocity dispersions \citep{wolf08}
should result in minimal errors. For WIYN objects, with $\sigma_{v}\sim110$
km $s^{-1}$, the measured velocity dispersions are used to smooth
the model stellar component, $ssp_{i\lambda}$. Narrow emission line
regions and sky residuals are uniquely masked for each object, giving
these regions no weight in the $\chi^{2}$ fitting. The broad components,
and in some cases entire broad lines, are left unmasked, though most
or all of the H-alpha region is masked out in many objects due to
significant sky residuals in the red end of the spectrum. Typically,
$\sim2200$ spectral data points are fit for Keck objects; $\sim1600$
spectral data points are fit for WIYN objects. Fig. \ref{fig3} depicts
the result of the method applied to 3C 273 4N. The results for all
other off-axis pointings are presented in the appendix, Fig. A1. Table
3 summarizes these findings.

\begin{figure*}
\centering{}\includegraphics[bb=65bp 60bp 540bp 710bp,clip,angle=90,width=14cm]{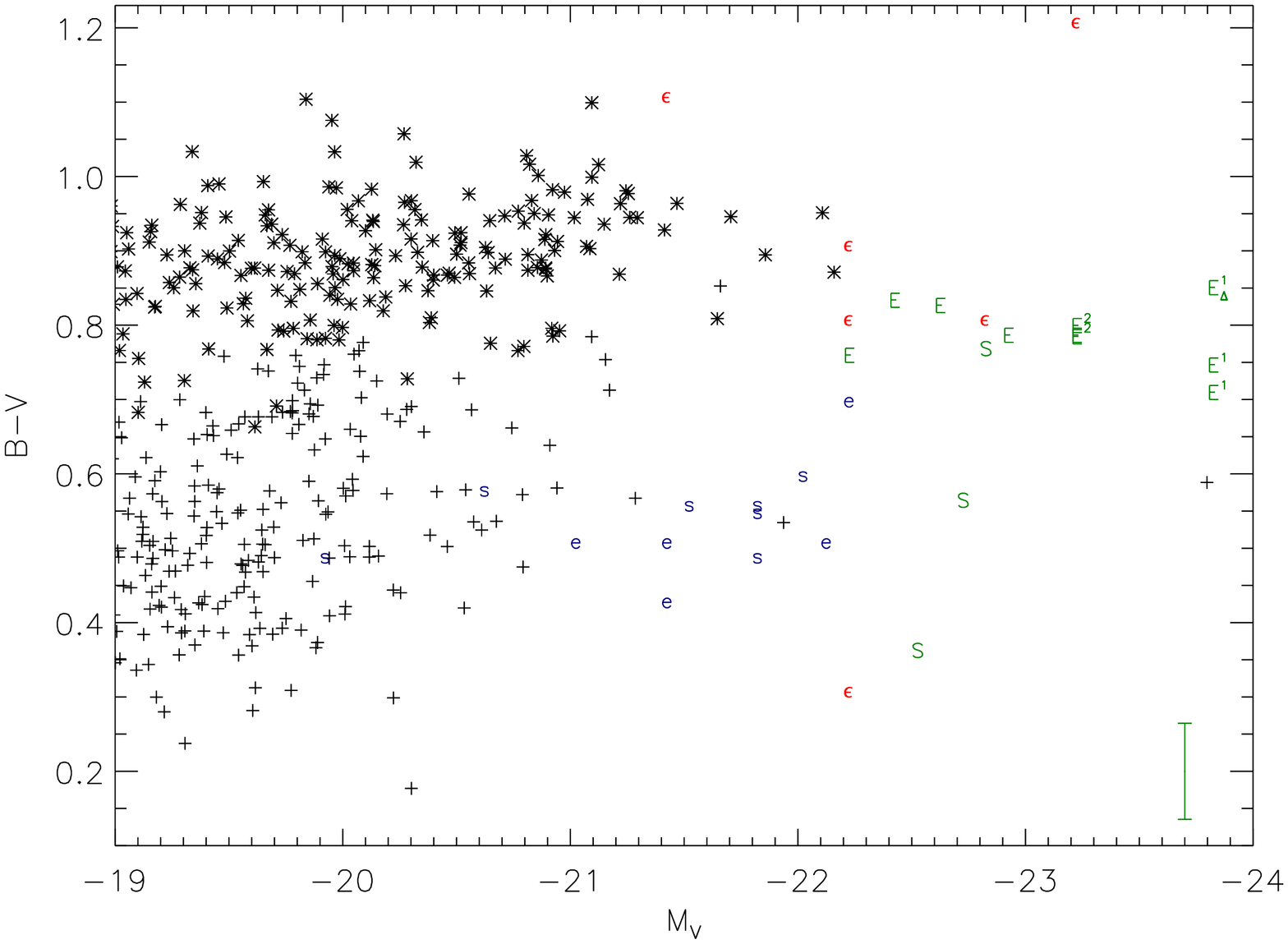}\caption{Rest frame colour-magnitude diagram for local quiescent galaxies and
active host galaxies. Black plus signs indicate (z$<$0.4) quiescent
late-type galaxies, while black asterisks indicate (z$<$0.4) quiescent
early-type galaxies \citep{wolf04}. J04 quasar hosts are blue lower
case letters ('s' for spiral or disc, 'e' for elliptical morphology).
H07 BL Lac hosts are all elliptical and are indicated by red '$\epsilon$'.
Quasar hosts from this work are indicated by green upper case letters
('S' for spiral or disc, 'E' for elliptical morphology). A typical
error bar is shown for our sample in the lower right. '1' and '2'
denote objects with multiple pointings, 4C 31.63 and PKS 2349-014
respectively. $\Delta$ denotes the observation with non-concurrent
on and off-axis observations. }

\end{figure*}

The reliability of these results are then determined with tailored
simulations. Synthetic spectra are generated with features representative
of the observed spectra by (1) degrading $S_{\lambda}$ to the observed
noise level, (2) applying the same mask constructed for the observed
spectrum, (3) using the appropriate observed on-axis spectrum as $Q{}_{\lambda}$
and (4) scaling the scattered quasar light to the observed contamination
level as determined by the model output%
\footnote{The contamination percentage is well recovered as tested by the Monte-Carlo
routines; for example, the contamination percentage for PKS 0736-017
is recovered within $\pm3.6$\%. This observation has the most limiting
signal-to-noise and provides an estimate of the error upper limit. %
}. An example of a simulation with the same on-axis spectrum, noise
characteristics, applied mask and percent scattered quasar light as
the observed off-axis spectrum of PKS 0736+017 is displayed in Fig.
5 top row. The bottom row of Fig. 5 presents the simulation results
for PG 1309+355. The PKS 0736+017 off-axis pointing has the worst
signal-to-noise, while the PG 1309+355 off-axis pointing has $\sim$average
signal-to-noise and quasar contamination. The results of all simulations
are shown in Table 3. Throughout the rest of the paper, mean offsets
are applied to model results and Monte-Carlo one sigma errors are
quoted.

\section{Comparison to recent studies}

We compare our spectroscopically derived rest frame \textit{B-V} host
colours to the available results from imaging studies. These imaging
studies spatially subtract the point spread function of the AGN to
reveal the host galaxy. Our method, which spectrally subtracts the
scattered quasar light, provides a complimentary technique to constrain
the host \textit{B-V} colours.

Our off-axis method probes a population of luminous quasars unattainable
by most other spectroscopic techniques. The stellar properties of
this population have yet to be determined for a large well defined
sample. This deficiency is due to the overwhelming emission from the
central quasar. Previous off-axis studies \citep[e.g.][]{hughes00}
have dealt with this problem by carefully selecting distant ($\sim$5$''$)
off-axis slit positions. This can significantly reduce the typical
scattered light observed off axis ($\sim$20\%). Our approach is to
more rigorously model the quasar light scattered off axis. This allows
for 1) scattered light fractions of up to $\sim$80\% and 2) more
freedom in slit positioning. Furthermore, our method draws from established
quiescent population synthesis models to more robustly model the stellar
content and to estimate the reliability of our results. 

Recent advances in on-axis observational techniques have allowed for
the study of luminous quasars \citep[][]{jahnke07,letawe07}. For
objects with $M_{V}(nuc)>M_{V}(host)$, our method can compliment
these studies by observing off-axis ($\gtrsim9kpc$) stellar content
with favorable signal-to-noise. We note that on-axis studies require
nucleus-to-host ratios of $\lesssim15$ to detect stellar absorption
features (L07). Objects such as NAB 0205+02, HE 0530-3755 and 3C 273
cannot be observed on-axis due to this criterion. Off-axis studies
do not have this limitation. Thus, our method is needed to obtain
optical stellar spectra for these objects.

In the following subsections we compare our study to previous imaging
and spectroscopic work in more detail. Furthermore, we quantify how
our sample compares to applicable SDSS AGN studies.

\subsection{Recent photometric studies}

Host \textit{B-V} colours have been reported for nearby ($z<0.197$)
quasars by \citet{jahnke04}, hereafter J04, and for nearby ($z<0.139$)
BL Lacs by \citet{hyvonen07}, hereafter H07. J04, studying multicolour
data from a sample of AGN distributed around the classical dividing
line in luminosity between Seyfert 1 galaxies and quasars, found bulge
dominated hosts to have bluer colours than expected compared to their
quiescent counterparts; disc hosts were also on average bluer, but
not by a significant amount. H07, also studying multicolour data,
found their exclusively elliptical BL Lac hosts to be on average bluer
than expected compared to their quiescent equivalent. Table 4 shows
the mean \textit{B-V} colours for the host galaxies in J04 and H07,
as well as in our sample and for quiescent galaxies of various morphologies
(from \citealt*{fukugita95}). In agreement with J04 and H07, we find
the rest frame \textit{B-V} colour for the elliptical host galaxies
of our sample to be bluer than one would expect given their quiescent
morphological counterparts. This provides confirmation that elliptical
quasar hosts are distinguishable in \textit{B-V} colour from inactive
elliptical galaxies using a spectroscopic technique. Spirals hosts
for our sample are found to be consistent with Sc galaxies which have
a \textit{B-V} colour of $\sim$0.54 \citep{fukugita95}. A comparison
to quiescent galaxies is also presented via a colour-magnitude diagram,
Fig. 6. Quiescent COMBO-17 \citep{wolf04} late-type galaxies (black
'$+$'), and quiescent COMBO-17 early-type galaxies (black '$*$')
are shown in relation to J04 hosts (blue lower-case letters; 's' for
spiral or disc, 'e' for elliptical morphology), H07 hosts (red '$\epsilon$'),
and this work (green upper-case letters; 'S' for spiral or disc, 'E'
for elliptical morphology). Quiescent early and late-type galaxy designations
are determined by the \textit{U-V} vs $M_{V}$ cut as proposed by
\citet{bell04} for z=0.25. A typical error bar is shown for our sample
in the lower right of Fig. 6.

Although we confirm bluer elliptical host galaxy colours, we find
that our sample is on average much redder than J04. This offset may
be due to the different effective radii probed since our method is
only sensitive to the stellar content observed off axis. For example,
if nuclear starbursts were a common occurrence in quasar hosts, then
one might expect the outer stellar populations to exhibit redder colours.
However, results from \citet{kauffmann03} suggest that the different
radii probed may not be responsible for this observed offset. Studying
a large sample of obscured AGN hosts, they find that star formation
in the hosts of powerful AGN (approximately one magnitude fainter
than a typical object in this study) is not concentrated primarily
in the nuclear regions; it is spread out over scales of at least several
kilo-parsecs. Furthermore, the host galaxies in the H07 study exhibit
a wide range of host colour gradients, with the majority displaying
bluer colours off axis. This negative colour gradient is also observed
for radio galaxies \citep{govoni00} and quiescent ellipticals \citep*{peletier90}.
Assuming these conclusions hold true for quasar hosts, one would expect
our sample to be biased toward bluer colours rather than the observed
red offset. 

An alternative explanation for our redder host colours compared to
the J04 study is that our sample is on average more radio loud%
\footnote{Radio luminosity is estimated utilizing NED and NVSS \citep{condon98}.
A spectral index of -0.5 is assumed. Upper limits are based on the
sensitivity of NVSS$~$ $\sim$ 2.3 mJy @ 1.4 GHz.%
}. The only J04 object which overlaps in \textit{B-V} colour with our
main sample is their most radio loud. The two objects from our sample
which overlap in colour with the main J04 sample are also our most
radio quiet. This relation of colour to radio luminosity is shown
in Fig. 7. Above $L_{5GHz}$ $\sim10^{40}\: erg\: s^{-1}$, the host
galaxies are systematically redder. Given the correlation between
radio luminosity and bulge mass found for our sample (Wolf \& Sheinis
2008, $L_{5GHz}\sim M_{bulge}^{3.56}$), $L_{5GHz}=10^{40}\: erg\: s^{-1}$
corresponds to $M_{bulge}\sim10^{11.5}M_{\odot}$. The rms scatter
of the correlation is 1.09 dex. This trend for radio loud objects
to be located in redder hosts is not surprising; \citet{best05} found
that radio-loud AGN are preferentially located in older more massive
galaxies. However the distinct transition we see in our sample should
be investigated on a larger sample. 

\begin{flushleft}
\begin{table}
\centering{}\begin{tabular}{cc}
Sample & \multicolumn{1}{c}{\textit{B - V}}\tabularnewline
\hline 
E Quasars (this work)  & 0.78$\pm$0.01\tabularnewline
S Quasars (this work) & 0.55$\pm$0.12\tabularnewline
E Quasars (J04) & 0.52$\pm$0.05\tabularnewline
S Quasars (J04) & 0.56$\pm$0.01\tabularnewline
Quasars $L_{5GHz}>10^{40}erg\: s^{-1}$(this work plus J04)  & 0.77$\pm$0.02\tabularnewline
Quasars $L_{5GHz}<10^{40}erg\: s^{-1}$(this work plus J04)  & 0.50$\pm$0.02\tabularnewline
E BL Lacs (H07) & 0.8$\pm$0.2\tabularnewline
E \citep{fukugita95} & 0.96\tabularnewline
Sb-Sc \citep{fukugita95} & 0.57\tabularnewline
Sc-Sd \citep{fukugita95} & 0.50\tabularnewline
\end{tabular}\caption{Available rest frame \textit{B-V} host mean colours and quiescent
mean colours. Reported errors are standard deviations of the mean.
'E' denotes elliptical and 'S' denotes spiral.}

\end{table}

\par\end{flushleft}

\begin{flushleft}
\begin{figure}
\centering{}\includegraphics[bb=60bp 175bp 520bp 660bp,clip,angle=90,scale=0.4]{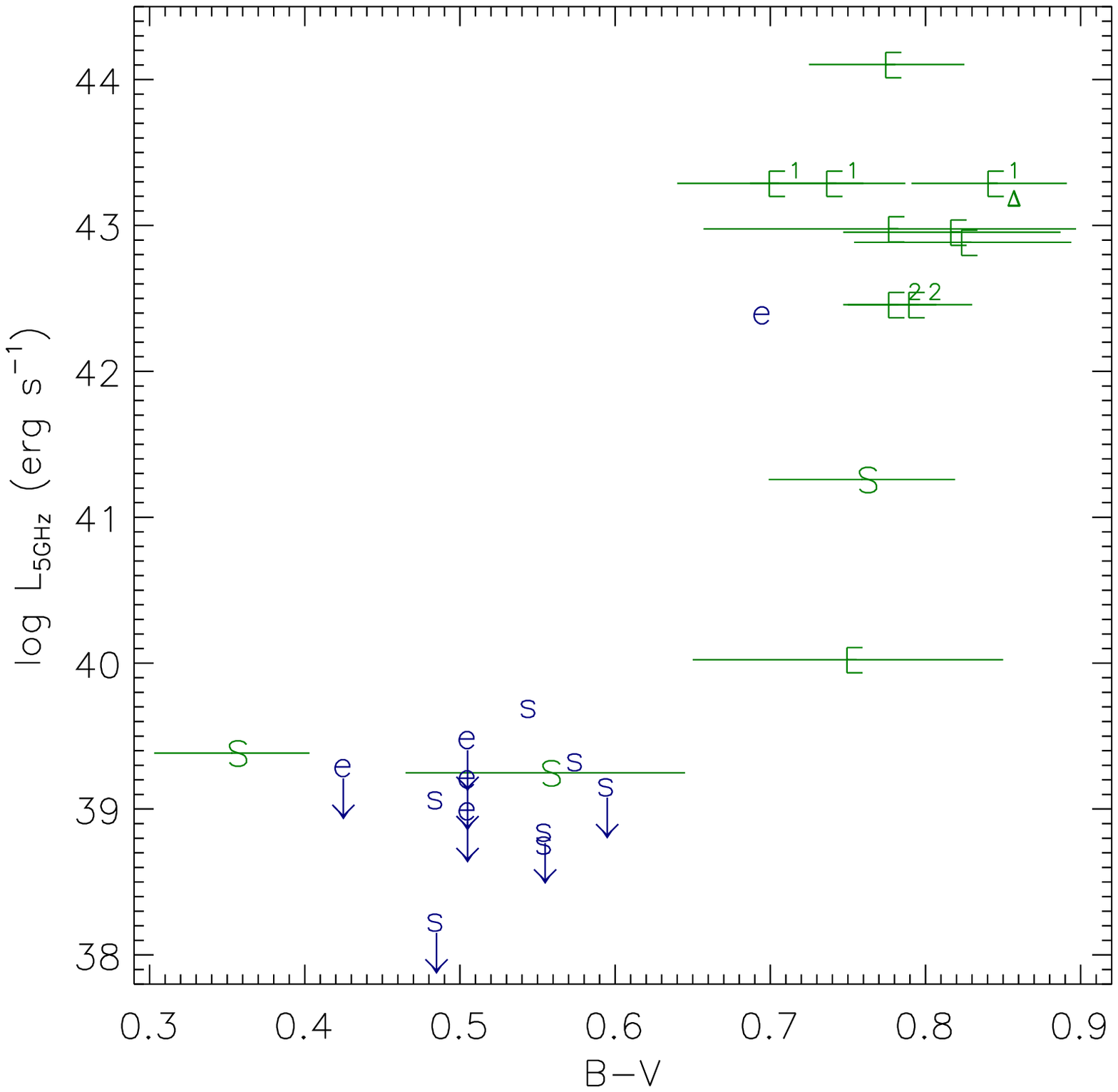}\caption{Radio luminosity as a function of rest frame \textit{B-V} colour.
Green upper case letters indicate the results of this study, blue
lower case letters indicate the results of J04. '1' and '2' shown
above denote objects with multiple pointings, 4C 31.63 and PKS 2349-014
respectively. $\Delta$ denotes the observation with non-concurrent
on and off-axis observations. H07's BL Lacs are not shown due to the
expected significant beaming of radio emission. }

\end{figure}

\par\end{flushleft}

\begin{flushleft}
\begin{figure}
\centering{}\includegraphics[bb=180bp 80bp 650bp 540bp,clip,angle=180,scale=0.4]{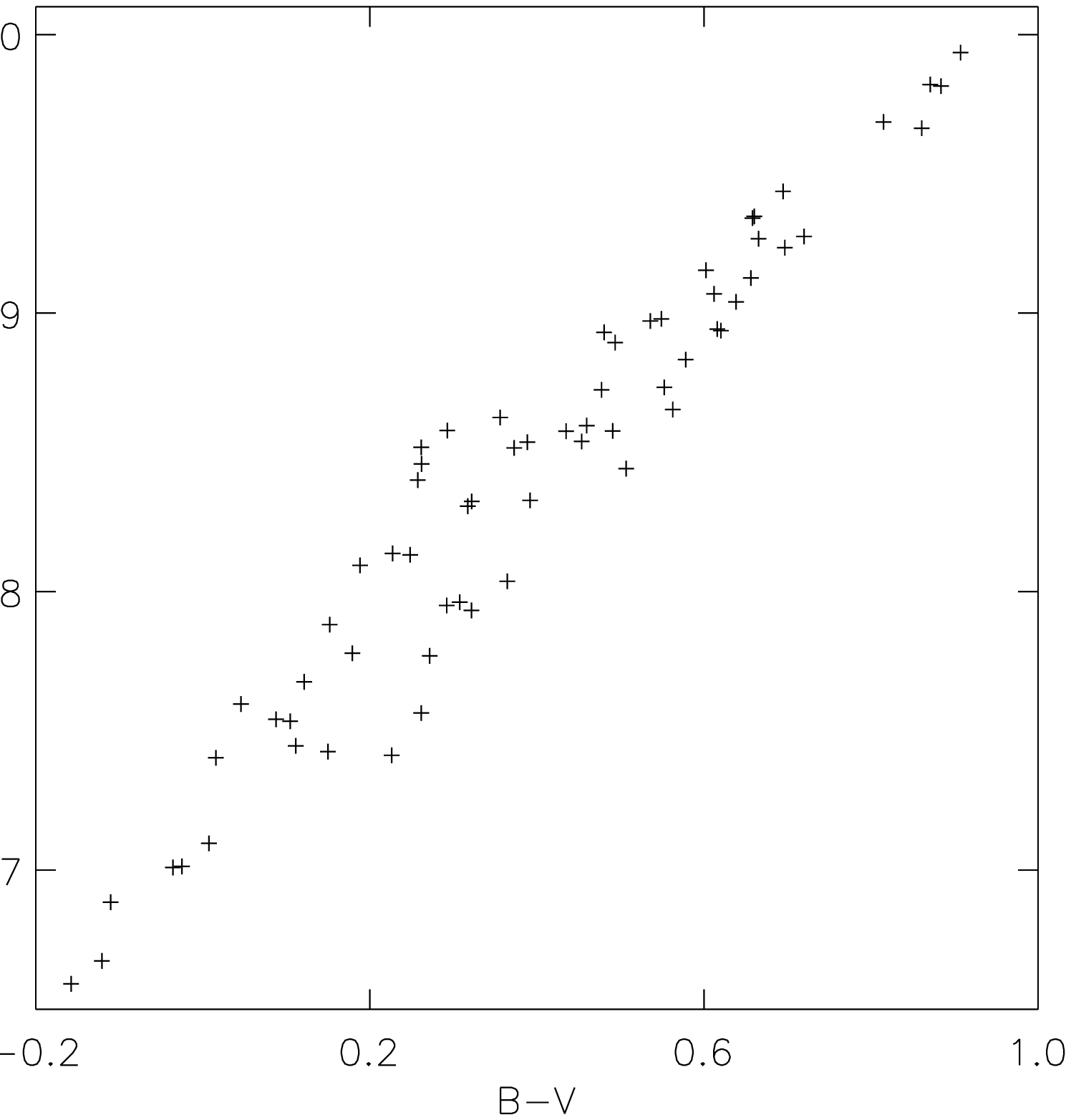}\caption{Relation between mean light-weighted stellar age and rest frame \textit{B-V}
colour derived from 65 randomly constructed stellar components, $\sum_{i=1}^{15}x_{i}\, ssp_{i\lambda}$. }

\end{figure}

\par\end{flushleft}

\begin{flushleft}
\begin{figure}
\centering{}\includegraphics[bb=50bp 190bp 510bp 655bp,clip,angle=90,scale=0.4]{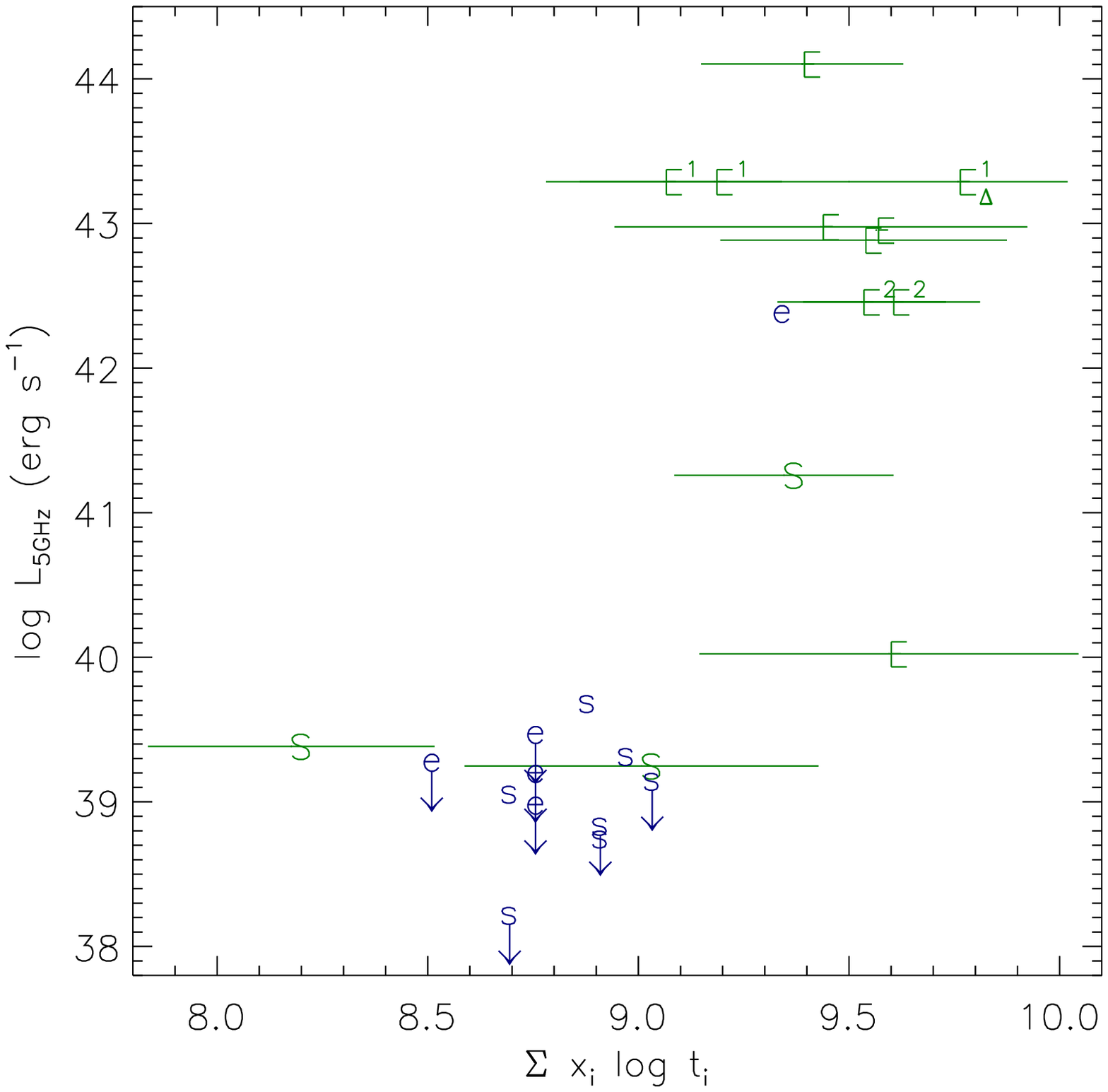}\caption{Radio luminosity as a function of mean stellar age weighted by flux.
Green upper case letters indicate the results of this study, blue
lower case letters indicate the results of J04. '1' and '2' shown
above denote objects with multiple pointings, 4C 31.63 and PKS 2349-014
respectively. $\Delta$ denotes the observation with non-concurrent
on and off-axis observations. }

\end{figure}

\par\end{flushleft}

How can this be understood in terms of stellar ages? The mean light-weighted
age is correlated to various colours \citep[e.g., \emph{u-r}, ][]{mateus06}.
We verify this correlation for our colour of interest, \textit{B-V},
by constructing 65 test spectra, Fig.8. Using this relation we convert
the \textit{B-V} colours of J04 to mean light-weighted ages, Fig.
9. While the observed colour trend is more pronounced, it is found
that all $\left\langle log\: t_{\star}\right\rangle _{L}$ values
above $L_{5GHz}=$$10^{40}\: erg\: s^{-1}$ are older than values
below this threshold. Previous studies (CF05; \citealt{mateus06})
have studied the mean light-weighted age and found this quantity to
be frequently associated with the most recent epoch of starbursts.
Assuming major merger progenitors for our quasar hosts with a significant
past starburst \citep[e.g., see][]{hopkins06}, $\left\langle log\: t_{\star}\right\rangle _{L}$
may indicate the approximate timing of the merger event. It should
be noted that our definition of $\left\langle log\: t_{\star}\right\rangle _{L}$
differs slightly from the definition adopted by CF05 and \citet{mateus06}
due to different SSP normalization conventions (see \S3.1). For our
sample, conversion to this alternative definition modifies our reported
$\left\langle log\: t_{\star}\right\rangle _{L}$ results by $\sim$1$\sigma$
or less.

To estimate this time-scale we divide the combined sample into two
groups; a radio bright sample with $L_{5GHz}>10^{40}erg\: s^{-1}$
displaying an average \textit{B-V} colour of 0.77 and a radio faint
sample with $L_{5GHz}<10^{40}erg\: s^{-1}$ displaying an average
\textit{B-V} colour of 0.50 (average colour for this grouping is also
reported in Table 4). $L_{5GHz}$ is found to correspond to bulge
mass and may provide a more significant division than morphology whose
classification is hampered by quasar point source subtraction and
un-modeled tidal features. We find that, on average, $\left\langle t_{\star}\right\rangle _{L}=2.7_{-0.5}^{+0.6}$
Gyrs%
\footnote{$\left\langle t_{\star}\right\rangle _{L}\equiv10^{\left\langle log\, t_{\star}\right\rangle }$%
} for the radio bright sample and $\left\langle t_{\star}\right\rangle _{L}=570_{-110}^{+140}$
Myrs for radio faint sample.

\subsection{Recent spectroscopic study with imaging}

This result for radio loud objects compares remarkably well to recent
findings of \citet{canalizo06} and \citet{bennert08}. \citet{canalizo06}
report preliminary deep Keck LRIS spectral analysis of luminous $z\sim0.2$
quasar hosts which finds evidence for major starburst episodes with
ages ranging from 0.6 Gyr to 2.2 Gyr. Incidentally at least two of
the host galaxies studied by Canalizo \& Stockton (in preparation)
overlap with our sample, PKS 0736+017 and PHL 909. They find that
both of these objects located in our radio bright group have massive
2.2 Gyr starbursts, consistent with our colour and mean light-weighted
age estimate. However, see Appendix A for an inconsistency in the
derived stellar populations for PHL 909. \citet{bennert08}, again
studying PKS 0736+017 and PHL 909 along with three other hosts in
deep HST ACS images, find shells and tidal tails indicative of merger
events. Consulting N-body simulations they conclude that in general
the observed fine structure can be explained by either a recent minor
merger or an older major merger $\sim$1Gyr ago. 

Implications of a $\sim$$ $Gyr poststarburst quasar are discussed
extensively in \citet{canalizo07} and \citet{bennert08}. In summary
the currently estimated (optically luminous) quasar lifetime ($10^{6}-10^{8}$yrs
\citep{yu02}) appears to be too short to be triggered at the same
time as the starburst. \citet{bennert08} conclude that this may imply
a scenario in which quasar activity is significantly delayed or a
scenario in which intermittent quasar activity encompasses a longer
duration. Alternatively, a more recent minor merger could explain
the observed quasar. Our results do not rule out this scenario. However,
the average mass-to-light ratio for our sample being $\sim$10 \citep{wolf08},
constrains any recent (t$\ll$1Gyr) starburst to be relatively insignificant
in mass.

\subsection{Off-axis studies}

The off-axis ($5$$''$) spectroscopic study of \citet[][hereafter H00]{hughes00}
consists of three matched subsamples of nine radio quiet quasars (RQQs),
ten radio loud%
\footnote{Radio loud in H00 is defined as $L_{5\, GHz}>10^{24}W\: Hz^{-1}\: sr^{-1}$,
which is roughly consistent with our adopted definition.%
} quasars (RLQs), and seven radio galaxies (RGs). Similar to our study,
all quasars are luminous ($M{}_{V}\lesssim-23$) and local ($0.1\leq z\leq0.3$). 

H00 determine the 4000 \AA\, break strength, an indicator of the
mean stellar age, for each object. The average values for RQQ hosts,
RLQ hosts, and RGs all indicate younger mean stellar ages than local
inactive elliptical galaxies, in agreement with our study. However,
H00 notes wide scatter in their measured values. Additionally, measurements
indicative of younger ages were typically associated with either poor
signal-to-noise data or with spectra that show significant scattered
quasar light. Thus H00 defer further analysis of the stellar populations
to the followup spectro-photometric modeling conducted by \citet[][hereafter N01]{nolan01}.

N01 develop two population synthesis models. The first model consists
of a 0.1 Gyr SSP and a second SSP component whose age and contribution
are allowed to vary to obtain the best $\chi^{2}$ fit. Like our study,
solar metallicity is assumed. However, two SSPs are used to construct
the stellar population in contrast to our model which incorporates
a base of 15 BC03 SSPs. N01 also develop a three component model which
uses the observed nuclear spectrum of the RQQ, 0054+144, to account
for scattered quasar light detected off axis. While our approach also
uses observed nuclear light to model off-axis scattered light, our
model does not assume a universal nuclear spectrum. The observed nuclear
spectrum is matched to the off-axis observation (e.g. the model input
for the quasar contamination in the 3C 273 off-axis spectrum is the
observed 3C 273 nuclear spectrum). Additionally, it should be noted
that whenever possible (9 out of 10 pointings) we have observed on
and off-axis observations simultaneously to avoid variation in the
quasar spectrum which could hamper an accurate scattered-light subtraction.
As a final point, we allow for the continuum shape of the nuclear
light observed off axis to be altered by the scattering efficiency
curve, which we have shown is necessary for reliable scattered light
removal \citep{sheinis02}. 

Interpretation of the N01 model output indicates that the stellar
mass of all three subsamples studied (RQQ, RLQ, and RG) is dominated
by old stars of age 8-14 Gyr. This result is used to support the claim
that quasar hosts are {}``to the first order, indistinguishable from
'normal' quiescent giant elliptical galaxies.'' If we take N01's
derived age to be representative of the mean mass-weighted stellar
age, then it is not clear that this quantity can be used to differentiate
between late and early type galaxies. \citet{mateus06}, studying
50,000 luminous galaxies from the SDSS, show that the distribution
of mean mass-weighted stellar ages is not bimodal, as both star-forming
and passive galaxies have formed a large fraction of their stellar
mass at early times ($\sim12.4$ Gyr in the past). We may comment
on the result of N01 by referring to the mass-to-light ratio derived
from the measured stellar velocity dispersion. \citet{wolf08} found
that the average mass-to-light ratio for our quasar host sample is
$\sim10$, which supports the dominance by mass of old stars. However,
as stated in \S5.1 our colour / $\left\langle log\: t_{\star}\right\rangle _{L}$
determination indicates that quasar hosts are bluer / younger than
quiescent ellipticals.

\subsection{On-axis studies}

J07 present a method that spatially subtracts the quasar point spread
function from the on-axis spectrum to extract the host galaxy light.
Unlike our method, which is constrained to collect light $\gtrsim9kpc$
off axis, this technique allows for the inner regions of the host
galaxy to be probed. Like our study, the J07 test sample consists
of nearby ($z<0.3$) quasars. However, this sample consists of quasars
that are typically less luminous in optical and radio bands than our
study. J07 model 13 host spectra with a two component BC03 SSP model.
Eight of the spectra were deemed trustworthy, yielding a typical light-weighted
stellar age of 1-2 Gyr. For a rough comparison, our sample average
$\left\langle t_{\star}\right\rangle _{L}$ is approximately 2 Gyr.
If we limit our sample to radio quiet objects, to more closely match
the sample of J07, we find an average $\left\langle t_{\star}\right\rangle _{L}$
of 1 Gyr. Confidence in this comparison is hampered by the small number
of objects probed and the sample mismatch: our targets are typically
more luminous. 

Despite this mismatch, there is one object that is modeled by both
groups, the radio loud quasar PKS 1302-102. While J07 find a stellar
age $<100Myr$, our results indicate a predominantly old, $>1Gyr$
stellar population. This disagreement may indicate a radial dependence
on the stellar populations or model systematics. 

L07, utilizing an on-axis method complementary to the J07 technique,
examine the host galaxies of 20 luminous quasars whose redshifts and
nuclear optical luminosities are well matched to our sample. However,
their optically selected sample contains far fewer RLQs. L07 measure
diagnostic absorption and emission lines and then compare these to
known quiescent galaxy values. Half of their sample was found to have
young Sc-like stellar populations. The sample examined by our study
indicates a redder and hence older stellar population than a typical
Sc galaxy. Our studies are again brought into closer agreement if
we consider only our RQQ hosts. However, no secure agreement can be
claimed due to our small RQQ sample. Further complicating the comparison,
L07 also model PKS 1302-102 and find evidence for a young Sc-like
stellar population. Additional interpretation of this disagreement
is discussed in Appendix A.

\subsection{Sloan Digital Sky Survey studies}

\begin{table}
\centering{}\begin{tabular}{cc}
Sample & \multicolumn{1}{c}{$\left\langle log\: t_{\star}\right\rangle _{L}$}\tabularnewline
\hline 
Quasars $L_{5GHz}>10^{40}erg\: s^{-1}$ (this work plus J04)  & 9.4$\pm$0.1\tabularnewline
Quasars $L_{5GHz}<10^{40}erg\: s^{-1}$ (this work plus J04) & 8.8$\pm$0.1\tabularnewline
Passive (M06) & 9.86\tabularnewline
AGN hosts (M06) & 9.66\tabularnewline
'green valley' (M06) & 9.53\tabularnewline
Star-forming (M06) & 8.91\tabularnewline
\end{tabular}\caption{Comparison of derived $\left\langle log\: t_{\star}\right\rangle _{L}$
values to those found in the SDSS study conducted by M06. Reported
errors are standard deviations of the mean.}

\end{table}

The Sloan Digital Sky Survey (SDSS) has been utilized to study AGN
hosts by a number of groups. These studies benefit from a large well
defined sample. Narrow-line AGN studies (\citealt[][hereafter K03]{kauffmann03};
\citealt[][hereafter M06]{zakamska03,mateus06}) and the broad-line
AGN study by \citet[][hereafter V06]{vandenberk06} are compared to
this work below.

K03 study the host galaxies of 22,623 local type 2 AGN. Modulo deficiencies
in AGN unification models, quasar hosts can be studied via their narrow-line
counter-parts. From 4000 \AA\, break measurements, K03 find that
high luminosity AGN have younger mean stellar ages than normal early-type
galaxies. Additionally, they find a large fraction of powerful type
2 AGN have experienced significant starbursts in the past 1-2 Gyr.
This is similar to our findings: $\left\langle t_{\star}\right\rangle _{L}=2.7_{-0.5}^{+0.6}$
Gyrs for the radio bright sample and $\left\langle t_{\star}\right\rangle _{L}=570_{-110}^{+140}$
Myrs for radio faint sample. However, there is little overlap in luminosity
between our samples. The K03 powerful (L{[}OIII{]}$\sim10^{41}erg\: s^{-1}$
) AGN are about one magnitude fainter than our sample's nuclear luminosities.
A smaller more luminous sample of SDSS narrow-line AGN has been examined
by \citet{zakamska03}. In qualitative agreement with K03, they too
find evidence for relatively blue / young host galaxies.

M06 divide a sample of 50,000 SDSS galaxies into star-forming, passive,
and AGN hosts spectral classes. The AGN probed by this study are comparable
in luminosity to the K03 sample. M06 find a clear bimodal distribution
for the mean light-weighted stellar age among star-forming and passive
galaxies. As summarized in Table 5, median values of $\left\langle log\: t_{\star}\right\rangle _{L}=8.91$
for star-forming galaxies and $\left\langle log\: t_{\star}\right\rangle _{L}=9.86$
for passive galaxies are found. The green valley is located at $\left\langle log\: t_{\star}\right\rangle _{L}\sim9.53$
and the median $\left\langle log\: t_{\star}\right\rangle _{L}$ for
AGN hosts is 9.66. The M06 median $\left\langle log\: t_{\star}\right\rangle _{L}$
for AGN hosts is heavily weighted toward the more numerous low luminosity
objects. In agreement with K03, they find that the typical $\left\langle log\: t_{\star}\right\rangle _{L}$
decreases for more luminous AGN. Our radio bright sample with $\left\langle log\: t_{\star}\right\rangle _{L}\sim9.4\pm0.1$
is in close proximity to the green valley (slightly younger and bluer).
We find the radio faint sample with $\left\langle log\: t_{\star}\right\rangle _{L}\sim8.8\pm0.1$
to fall very close to the peak of star-forming distribution. Qualitatively
we find this to be in overall agreement with our colour-magnitude
diagram, Fig. 6. 

V06 study the host galaxies of 4,666 local broad-line AGN. Like K03
and M06, they find that higher luminosity AGN hosts are bluer /younger
than normal early-type galaxies. This sample contains $\sim1750$
quasars with luminosities comparable to our study. However, the model
employed by V06 fails when the on-axis host flux fraction falls below
10\% or when the signal-to-noise is less than 10. These criteria prevent
the study of $\sim70\%$ of their luminous objects and would exclude
the majority of the objects (7 of 10) in this study.

\section{Summary and Conclusions}

We have presented an off-axis technique to spectroscopically constrain
the colour and the stellar ages of quasar ($M_{V}(nuc)<-23$) host
galaxies. Our method draws heavily from the quiescent galaxy model
of CF05, utilizing a basis of BC03 SSPs and \citet{cardelli1989}
dust extinction in a similar fashion. Complicating our $\chi^{2}$
fitting routine is the residual scattered quasar light (a combination
of atmospheric, instrumental and host galaxy scattered light) which
must be accounted for in the off-axis spectrum. Scattered light is
modeled by altering the the observed nuclear spectrum by a low order
polynomial, while simultaneously fitting the constituent stellar populations
of the host galaxy. Furthermore, Monte-Carlo simulations are tailored
to each observed pointing. Thus, the ability of the model to recover
known parameters from synthetic spectra is determined. It is found
that quasar host \textit{B-V} colour and $\left\langle log\: t_{\star}\right\rangle _{L}$
are well recovered. These parameters are then compared to previous
studies. Overall consensus is found giving further credence to our
model (however, see Appendix A for discussion of individual cases
that potentially disagree with previous results).

Our method probes a population of luminous quasars unattainable by
most other techniques. The stellar properties of this population have
yet to be determined for a large well defined sample. For the smaller
sample ($N_{obj}\sim20$) studies that can access this luminous regime,
our method can be used in conjunction to gain further insight. For
example, our technique can observe the off-axis ($R_{obs}$$\gtrsim9kpc$)
stellar content of objects (nucleus-to-host ratio$\gtrsim$15) inaccessible
to the on-axis studies of J07 and L07. 

The full potential of this method has yet to be realized. To date,
we have analyzed the properties of 10 quasar host galaxies. With this
sample, we confirm that elliptical quasar hosts are distinguishable
(bluer) from inactive ellipticals in rest frame \textit{B-V} colour.
Additionally, we note a trend for radio luminous quasars to be located
in redder host galaxies in comparison to their less radio luminous
counterparts. However, general conclusions await further observations.
Currently we have 10 additional WIYN off-axis observations in the
process of being reduced. Furthermore, time has been allotted on the
WIYN telescope to increase our total analyzed sample size to $\sim25$
objects. This sample will consist of roughly an equal number of RLQs
and RQQs and allow for 1) more rigorous comparison to other studies
due to the increase in sample size, 2) further verification of our
technique with the targeting of L07 overlap objects (specifically
HE 0914-0031 and HE 0956-0720), 3) investigation of the trends found
in our preliminary sample. In the near future, the Robert Stobie Spectrograph
(RSS) on the 11-m Southern African Large Telescope (SALT) will be
employed in our ongoing observational campaign. This phase of the
project will increase the observable sample by its location in the
Southern hemisphere and by allowing for higher redshift objects to
be targeted ($z\sim0.5$). 

Future work will build on the current model to analyze the narrow-line
emission. This will provide insight into the star formation rate,
the emission mechanism, the metallicity, and the stellar content (e.g.
see L07). The accomplishment of this program will provide important
observational constraints on luminous quasars.

\section*{Acknowledgements}

The authors would like to thank Joseph Miller, Roberto Cid Fernandes,
and Laura Trouille for insightful conversations on various aspects
of this work. We are also in debt to Amy Barger for use of computing
resources. Some of the data presented herein were obtained at the
W. M. Keck Observatory, which is operated as a scientific partnership
among the California Institute of Technology, the University of California,
and the National Aeronautics and Space Administration. The Observatory
was made possible by the generous financial support of the W. M. Keck
foundation. Data were also obtained at the WIYN Observatory, which
is a joint facility of the University of Wisconsin-Madison, Indiana
University, Yale University, and the National Optical Astronomy Observatories.

\appendix{}

\section{Comments on individual hosts}

In this section, we describe the stellar populations found for individual
hosts and compare these to previous studies. Objects are listed in
order of greatest 5 GHz luminosity to least, as in Table 1. Unless
otherwise mentioned: young $\equiv$ (t $<$ 100 Myr), intermediate
$\equiv$ (100 Myr \ensuremath{\le} t \ensuremath{\le} 1 Gyr), and
old $\equiv$ (t $>$ 1 Gyr), where t is the age of the starburst.
Additionally, stellar populations (young, intermediate, and old) are
reported by percent flux contribution. Results from the Monte-Carlo
simulations, namely the mean offset and the one sigma errors, are
utilized in the following comments. By convention, application of
offsets may not force the percent flux contribution to be negative
or greater than 100\%. For output parameters without offsets applied
see Table 3. Fig. 4 and Fig. A1 are also shown without offsets applied.
As mentioned in \S3.2.3, error estimates reflect one sigma performance;
thus we expect statistically 4 of our 13 pointings to lie beyond our
error limits.

\begin{figure*}
\begin{centering}
\includegraphics[bb=55bp 110bp 750bp 520bp,clip,angle=90,width=14cm]{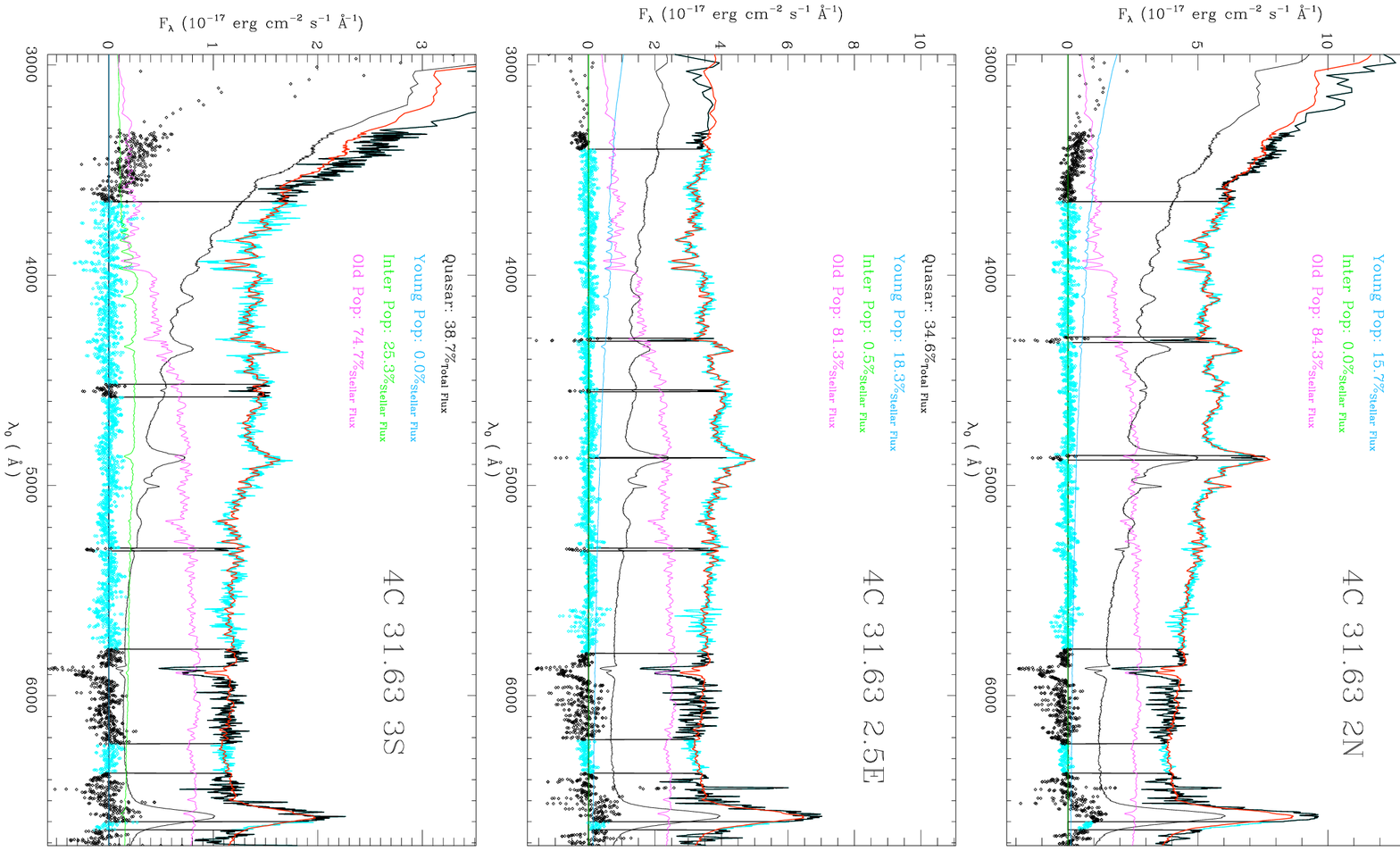}
\par\end{centering}

\caption{Model fit to the observed off-axis spectra. The meanings of the symbols
and colours are the same as in Fig. 4.}

\end{figure*}
\addtocounter{figure}{-1}  

\begin{figure*}
\begin{centering}
\includegraphics[bb=55bp 110bp 750bp 520bp,clip,angle=90,width=14cm]{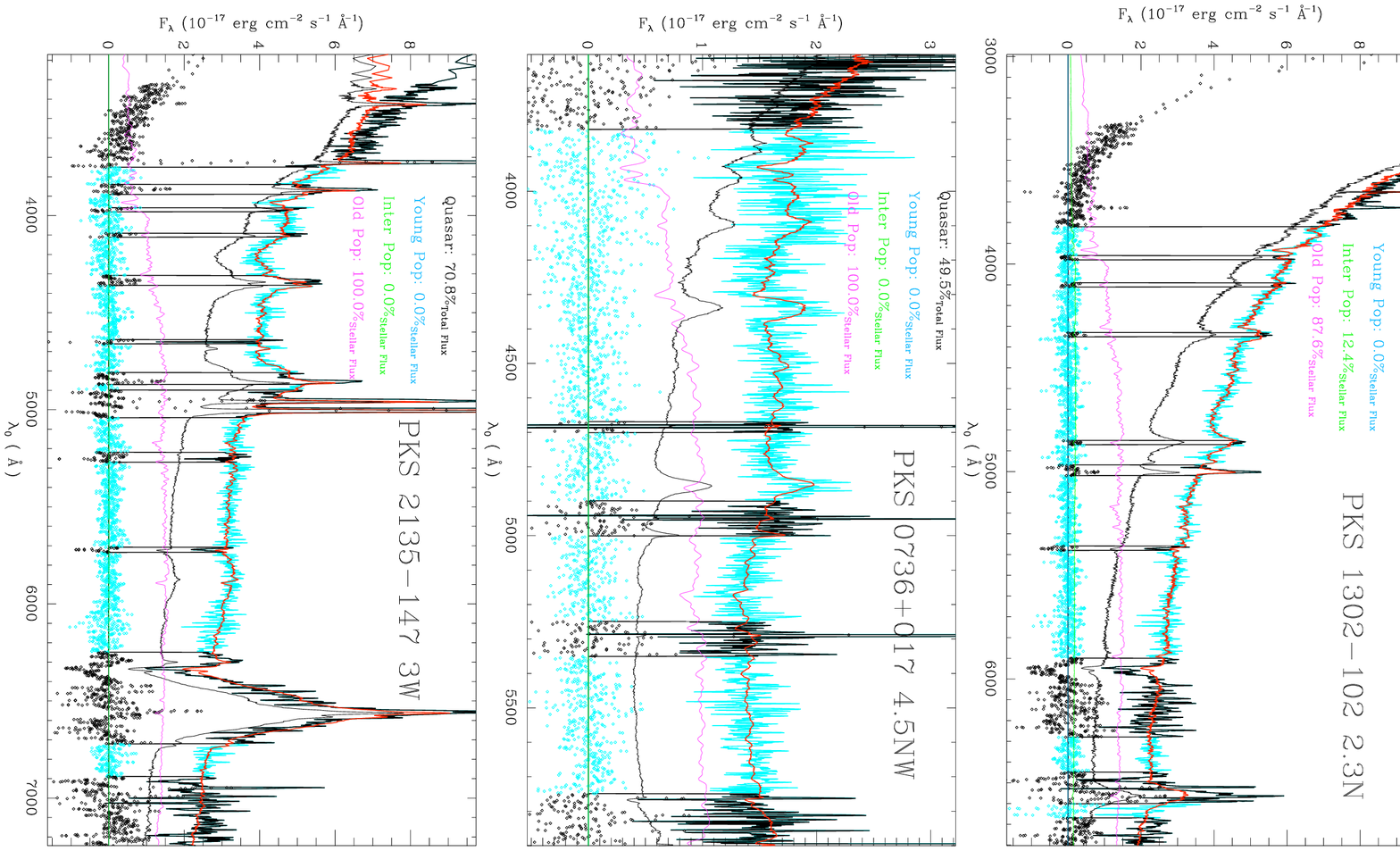}
\par\end{centering}

\caption{continued}

\end{figure*}
\addtocounter{figure}{-1}  

\begin{figure*}
\centering{}\includegraphics[bb=55bp 106bp 750bp 520bp,clip,angle=90,width=14cm]{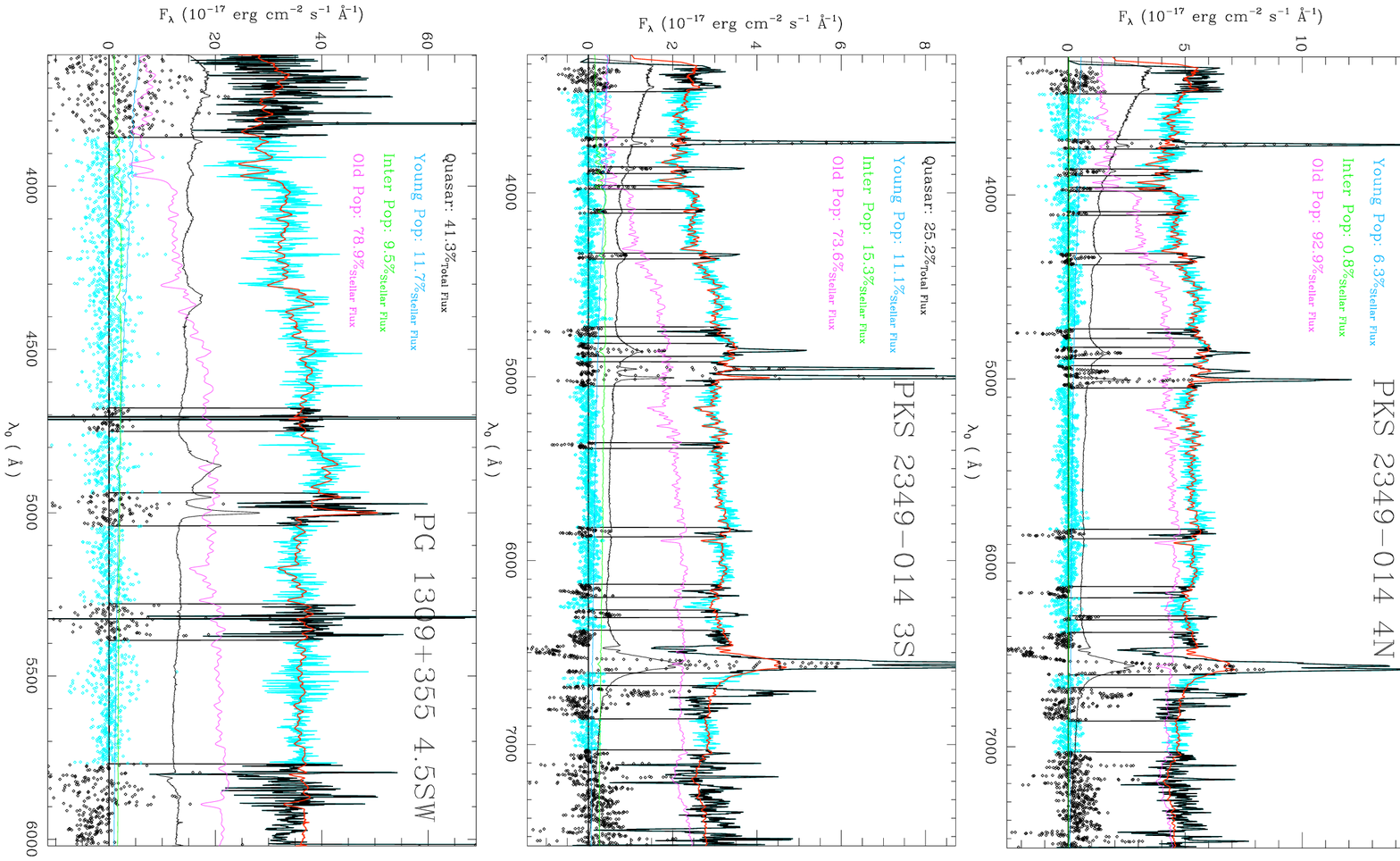}\caption{continued}

\end{figure*}
\addtocounter{figure}{-1}  

\begin{figure*}
\centering{}\includegraphics[bb=55bp 103bp 750bp 520bp,clip,angle=90,width=14cm]{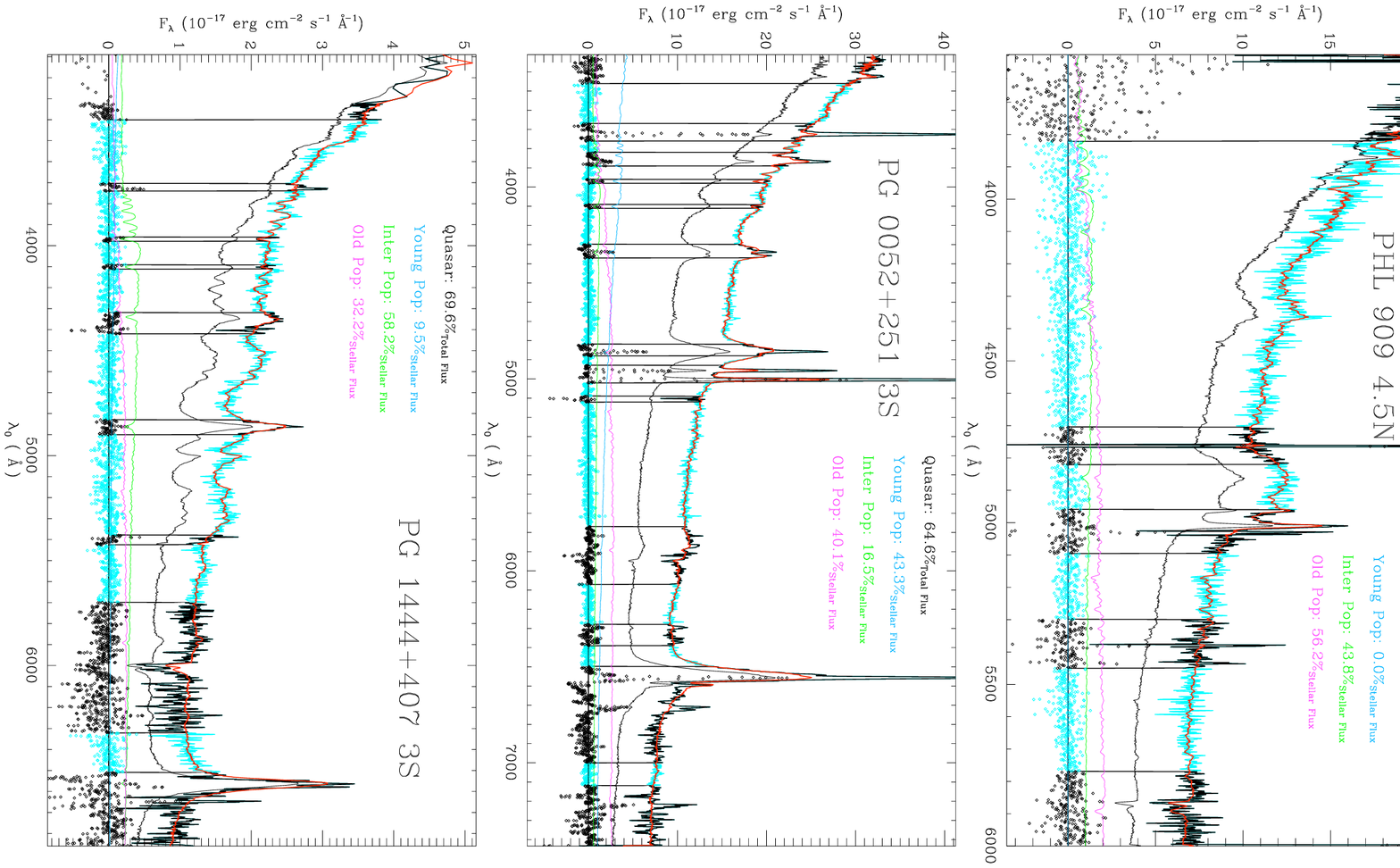}\caption{continued}

\end{figure*}

\smallskip{}

\textbf{\textit{3C 273}}\textit{ }(PG 1226+023) -- The host galaxy
is classified as an elliptical galaxy via HST imaging \citep{bahcall97, hamilton02}.
The off-axis spectral analysis of \citet{boroson85} qualitatively
found evidence for a significant contribution from an old stellar
population. Our results, 88.8$\pm$13.8 \% flux contribution from
an old stellar population at an observed radius of 11.8 kpc, are consistent
with these studies. Somewhat below the 1$\sigma$ error, a young stellar
component is found by the routine. Residual OII and OIII narrow line
emission is noted.

\textbf{\textit{4C 31.63}} (Q2201+315) -- 4C 31.63 is a radio loud
quasar whose host galaxy has been classified as an elliptical by HST
imaging \citep{bahcall97, hamilton02} and by near-infrared adaptive
optics imaging \citep{guyon06} . Our results for the Northern and
Eastern pointings are consistent with a predominantly old stellar
population. Toward the south we find evidence for an intermediate
stellar population not found in the other pointings. Note that the
3S pointing has non-concurrent on and off-axis observations, see \S
4 for details. No narrow line emission is found for this object.

\textbf{\textit{PKS 1302-102}} (HE 1302-1017) -- The host galaxy
of this radio loud quasar has been classified as a disturbed elliptical
by various studies \citep{hutchings92, bahcall97, hamilton02, guyon06}.
\citet{jahnke07} and \citet{letawe07} have found evidence from on-axis
spectroscopy for a young spiral-like stellar population. \citet{jahnke07}
note that there are almost no stellar absorption lines with the exception
of a weak Ca $II$ K and Mg $I$ line. This is not the case in our
observed off-axis spectrum; Ca $II$ K $ $$\lambda3933$ and H $\lambda3969$
are prominent with clear Mg $I$ b $\lambda5175$ and Na D $\lambda5895$
absorption features. The stellar component is found to be best fit
by a 88.9$\pm$14.8\% old population with a statistically insignificant
intermediate population. \citet{jahnke07} and \citet{letawe07} analyse
the stellar content with on-axis spectra, potentially probing much
closer in to the central SMBH. Thus a plausible explanation for this
disagreement is the different effective radii probed by our studies;
however, other systematics are not ruled out. In our observed off-axis
spectrum, scattered light from the central quasar is significant $~$$\sim$68.4\%,
effectively decreasing the signal-to-noise of the stellar component.
Additionally, the Balmer decrement has been masked due to nebular
emission, potentially removing important spectral information. On
the other hand, we estimate the nucleus-to-host ratio to be 13.5,
which is relatively close to the stated upper limit of $\sim15$ to
reliably detect absorption features with the on-axis method (L07).
Regardless, the observed prominent Ca H and K absorption lines support
the claim that a significant old population is present at the radius
probed $\sim13$ Kpc. Residual OII and OIII narrow line emission is
noted.

\textbf{\textit{PKS 0736+017}} -- The host galaxy has been classified
as elliptical (\citealt*{wright98} \citealt{ mclure99, falomo00, hamilton02}).
Others describe the host as a disturbed elliptical based on near-infrared
imaging \citep{dunlop93} and deep HST imaging \citep{bennert08}.
Off-axis spectroscopic studies have found a stellar component indicative
of a 12 Gyr population \citep{hughes00, nolan01}. Canalizo \& Stockton
(in preparation) find a 2.2 Gyr starburst in this host \citep{bennert08}
which falls in our 'old' age bin. Our results indicate 100$\pm$12.4\%
flux contribution due to an old population. No narrow line emission
is found for this object.

\textbf{\textit{PKS 2135-147}} -- The host galaxy of this radio loud
quasar has been classified as elliptical by \citet{bahcall97} and
\citet{hamilton02}. Our results are consistent with the expected
stellar population of an elliptical galaxy. We find 100$\pm$12.0\%
old stellar population at a radius of approximately 12.5 Kpc. Multiple
residual narrow emission lines are noted for this object.

\textbf{\textit{PKS 2349-014}} -- The host galaxy of this radio loud
quasar has been classified as a highly disturbed elliptical by many
imaging studies \citep{bahcall97, guyon06}. An off-axis spectroscopic
study by \citet{nolan01} found a dominant old stellar population
of 12 Gyr with a low level young population. Off-axis studies \citep{miller03, wolf08},
utilizing the same data analysed in this paper, have also found evidence
for a predominately old stellar population with a lesser young population.
Our results are consistent with a predominantly old population by
flux. The northern pointing (14.12 kpc), probing large tidal arms,
is consistent with an entirely old stellar population; whereas the
closer-in Southern pointing (9.47 kpc) we find evidence for a small
intermediate population (16.1$\pm13.2\%$). Multiple residual narrow
emission lines are noted for this object.

\textbf{\textit{PG 1309+355}} -- HST imaging studies classify this
object as a spiral \citep{bahcall97, hamilton02}, while the near-infrared
study of \citet{guyon06} classifies the host as an elongated elliptical.
If this object is truly a spiral, then this is a very rare object
displaying a large radio luminosity, just below our radio-loud threshold.
Our results indicate a predominantly old population. No significant
narrow emission lines are noted.

\textbf{\textit{PHL 909}} (0054+144) -- The host is an elliptical
galaxy as classified by the imaging studies of \citet{bahcall97}
and \citet{hamilton02} . The near-infrared imaging of \citet{dunlop93}
detects extended emission toward a companion. Recent deep HST imaging
\citep{bennert08} detects tidal tails and shells. \citet{barthel06},
analysing the far-infrared / radio correlation, finds evidence of
recent or on going star formation. Off-axis spectroscopic studies
have either classified the stellar population as old \citep{nolan01}
or have reported ambiguous results \citep{hughes00}. Canalizo \&
Stockton (in prep.), analysing $\sim$2 hr exposure Keck LRIS data,
find a 12 Gyr population combined with a 2.2 Gyr starburst \citep[as reported in][]{canalizo06}.
This would correspond to a 100\% old population given our adopted
age bins; however our results indicate 59.6$\pm$12.4\% old and 48.7$\pm$20.0\%
intermediate by flux. This pointing has the highest scattered quasar
component at 77.8\% by flux. The scattered light contamination is
highest at the blue end of the spectrum where the 4000\AA ~break
and various stellar absorption features are located. Thus our Monte-Carlo
error simulations, which normally are not constrained to have steep
blue scattering efficiency curves, may be under estimating the actual
error. To test this scenario, we performed Monte-Carlo simulations
with input parameters indicative of PHL 909 and constrained to have
a steep scattering efficiency curve, as found by the best fit model
output. The error estimates and offsets were not significantly affected
by this alteration%
\footnote{As listed in Table 1 and Table 2, the original error simulations found
the following offsets and $\pm1\sigma$ errors: $-8.3\pm18.4\%$ for
young, $+4.9\pm20.0\%$ for intermediate, $+3.4\pm12.4\%$ for old,
$+0.02\pm0.10$ for \textit{B-V} colour, and $+0.14\pm0.45$ for $\left\langle log\: t_{\star}\right\rangle _{L}$.
The steep blue scattering simulations found: $-3.1\pm22.0\%$ for
young, $-1.8\pm22.5\%$ for intermediate, $+4.9\pm11.9\%$ for old,
$+0.01\pm0.08$ for \textit{B-V} colour, and $+0.03\pm0.45$ for $\left\langle log\: t_{\star}\right\rangle _{L}$.%
}. If our observations are co-spatial with those of Canalizo \& Stockton,
then the difference in the results may simply be a $3\sigma$ outlier
arising from separate observations and different analysis methods.
Regardless, construction of a solar BC03 stellar spectrum consistent
with the findings of \citet{canalizo06} results in values for the
mean light-weighted age and \textit{B-V} colour that are consistent
(within 1$\sigma$) of our findings. Therefore, we believe that the
main discussion regarding the more reliably recovered parameters,
colour and $\left\langle log\: t_{\star}\right\rangle _{L}$, are
robust. 

\textbf{\textit{PG 0052+251}} -- The host galaxy associated with
this radio quiet quasar is a spiral as classified by various imaging
studies \citep{bahcall97, hamilton02}. Our results are consistent
with a significant young population of 35.3$\pm$19.2\% by flux and
an old population of 43.7$\pm11.5$\%. Multiple residual narrow emission
lines are noted for this \nopagebreak object.

\textbf{\textit{PG 1444+407}} -- The host of this radio quiet quasar
is classified as a spiral by \citet{hamilton02}. \citet{bahcall97}
note that the host has the appearance of an elliptical; however the
light profile is best fit by an exponential disc model. Additionally,
\citet{ho05} estimate a significant star formation rate (19.4 $M_{\odot}yr^{-1}$)
based on the {[}O II{]} luminosity. Our results indicate a 60.8$\pm$27.6\%
intermediate population, a 37.4$\pm$17.7\% old population and a statistically
insignificant young population. Multiple residual narrow emission
lines are noted for this object.
\bibliographystyle{mnras}

\end{document}